\RequirePackage{fix-cm}
\documentclass[twocolumn,epjc3]{svjour3}
\smartqed  
\RequirePackage{graphicx}
\journalname{Eur. Phys. J. C 81 (2021) 635, https://doi.org/10.1140/epjc/s10052-021-09457-2}

\usepackage{mathrsfs}
\usepackage{times}
\usepackage{amsmath}
\usepackage{graphicx}
\usepackage{sidecap}
\usepackage{txfonts}
\usepackage{color}
\usepackage{ulem}
\usepackage[colorlinks=true,citecolor=blue,linkcolor=magenta]{hyperref}
\usepackage{subeqnarray}
\usepackage{verbatim}
\usepackage{epstopdf}
\usepackage{cases}
\usepackage[symbol]{footmisc}
\usepackage{subfigure}
\usepackage{epsfig}
\usepackage[titles,subfigure]{tocloft} 
\usepackage{longtable}
\usepackage{widetext}

\begin{document}

\title{Boer-Mulders function of the pion and pretzelosity distribution of the proton  in the polarized pion-proton Drell-Yan process at COMPASS}


\author{Xiaonan Liu\thanksref{addr1}
        \and
        Bo-Qiang Ma\thanksref{e2,addr1,addr2,addr3} 
}

\thankstext{e2}{mabq@pku.edu.cn}


\institute{School of Physics and State Key Laboratory of Nuclear Physics and Technology, Peking University, Beijing 100871, China \label{addr1}
           \and
           Center for High Energy Physics, Peking University, Beijing 100871, China \label{addr3}
            \and
           Collaborative Innovation Center of Quantum Matter, Beijing, China \label{addr2}
}

\date{Received: date / Accepted: date}

\maketitle

\begin{abstract}
We present a phenomenological analysis of the $q_{\mathrm{T}}$-weighted transverse spin dependent azimuthal asymmetry recently measured by the COMPASS Collaboration in polarized pion-proton Drell-Yan process. In the kinematical regimes explored by experiments, we consider the leading-twist contributions from the Boer-Mulders distribution functions $h_{1}^{\perp}(x,k_{\mathrm{T}}^{2})$ of both the pion and the proton, the transversity distribution $h_{1}(x,k_{\mathrm{T}}^{2})$ and the pretzelosity distribution $h_{1\mathrm{T}}^{\perp}(x,k_{\mathrm{T}}^{2})$ of the proton, with the unpolarized transverse-momentum-dependent distribution function $f_{1}(x,k_{\mathrm{T}}^{2})$ of the proton being also involved in the calculation. By comparing the data reported by the COMPASS Collaboration with theoretical calculated results, we find that the COMPASS measurements represent the first experimental evidence of the Boer-Mulders effect in polarized Drell-Yan process. We also test the universality between proton and pion Boer-Mulders functions.

\end{abstract}

\section{Introduction}
\label{sec_intro}
%

Processes like semi-inclusive deep-inelastic lepton nucleon scattering~(SIDIS) and the Drell-Yan~(DY) process factorize~\cite{Collins:2011zzd} at leading twist allowing to access information on transverse-momentum-dependent~(TMD) parton distribution functions~(PDFs). TMD PDFs encode all possible correlations between the nucleon spin, the parton spin and the transverse component of the intrinsic parton momentum, $k_{\mathrm{T}}$, and thus allow for a three-dimensional ``3-D'' representation of the nucleon structure in momentum space. Each TMD PDF enters the SIDIS and DY cross sections with a specific dependence on azimuthal angles defined by the kinematics of the process. Hence, it can be extracted through measurements of azimuthal~(spin) asymmetries, which providing crucial information on spin-orbit correlations and the orbital angular momentum of confined quarks inside the nucleon. The advantage for extraction of PDFs of the Drell-Yan process~\cite{Drell:1970wh}, i.e., massive lepton-pair production in hadron-nucleon collisions~($hN\rightarrow l\bar{l}X$), is that there is no need for any fragmentation functions. Hence, it makes Drell-Yan processes unique and powerful to provide direct access to parton distribution functions. In this paper, we focus our attentions on the transverse spin dependent azimuthal asymmetries~(TSAs) arising in Drell-Yan cross sections.

These types of TSAs arise from the coupling of two different chiral-odd parton distributions. The key ingredient for these TSAs is the Boer-Mulders function $h_{1}^{\perp}$~\cite{Boer:1997nt}, which is a naively time-reversal odd transverse-momentum-dependent distribution function and provides the necessary phase required for TSAs. The Boer-Mulders function $h_{1}^{\perp}(x,k_{\mathrm{T}}^{2})$ describes a correlation between the transverse spin and the transverse momentum of a quark inside an unpolarized hadron. It is important to note that sensitivity to the sign of the Boer-Mulders function exists only in the polarized Drell-Yan process. For the unpolarized Drell-Yan process the azimuthal $\cos 2\phi$ asymmetry contains a convolution of two Boer-Mulders functions, while in the polarized DY process the Boer-Mulders function appears also coupled to other TMD PDFs, e.g. transversity distribution functions $h_{1}$~\cite{Ralston:1979ys}. The coupling $h_{1}^{\perp}\otimes h_{1}$ was first introduced and analyzed in Ref.~\cite{Boer:1999mm} as an alternative mechanism for TSA and a method of accessing the transversity distribution functions $h_{1}$~\cite{Ralston:1979ys,Jaffe:1991kp}.

Therefore, measuring TSAs gives access to a ratio of convolutions of TMD PDFs. The convolutions are usually solved assuming a certain $k_{\mathrm{T}}$-dependence of TMD PDFs (e.g., a Gaussian parametrization). The weighted TSAs have recently received increasing attention because of the de-convolution of TMD PDFs. The weighted TSAs, on the other hand, represent a way to avoid making any assumption on the $k_{\mathrm{T}}$-dependence. They take advantage of the fact that, if one integrates the structure functions over $\vec{q}_{\mathrm{T}}=\vec{k}_{a\mathrm{T}} + \vec{k}_{b\mathrm{T}}$ with appropriate weights $W_{X},$ the convolutions can be easily solved. For details of the $q_\mathrm{T}$-weighted TSAs see Section~\ref{sec_model}.

Recently, COMPASS published the first measurement of the transverse spin dependent azimuthal asymmetries in the polarized Drell-Yan process, ($\pi^{-}p^{\uparrow}\rightarrow \mu^{-}\mu^{+}X$), induced by a $190~\mathrm{GeV/c}$ $\pi^{-}$ beam scattering off a transversely polarized $\mathrm{NH_{3}}$ target~\cite{Longo:2019bih}. The measurement of the Sivers and other azimuthal asymmetries at the same hard scale in polarized SIDIS and Drell-Yan provides a unique possibility to test prediction in QCD universal features of transverse momentum dependent parton distribution functions. More recently, a lot of theoretical studies and phenomenological analysis on TMD PDFs by $q_{\mathrm{T}}$-weighted TSAs have been carried out. In the $q_{\mathrm{T}}$-weighted framework, a point-by-point method has been adopted to extract the Sivers function $f_{1\mathrm{T}}^{\perp}(x,k_{\mathrm{T}}^{2})$~\cite{Alexeev:2018zvl}, which describes the influence of the transverse spin of the nucleon onto the quark transverse momentum distribution. The purpose of the present work is to perform a phenomenological analysis of the recent TSA measurements in order to extract some information about the Boer-Mulders function, transversity and pretzelosity distribution functions, and to check their universality.

In this work, we calculate the $q_{\mathrm{T}}$-weighted Boer-Mulders-transversity asymmetry $A_{\mathrm{T}}^{\sin \left(2 \varphi-\varphi_{S}\right) q_{\mathrm{T}}/{M_{\pi}}}$ and Boer-Mulders-pretzelosity asymmetry $A_{\mathrm{T}}^{\sin \left(2 \varphi+\varphi_{S}\right) q_{\mathrm{T}}^{3}/{2M_{\pi}M_{P}^{2}}}$. As we will see in Section~\ref{sec_model}, the two asymmetries are sensitive to the Boer-Mulders functions of the pion, convoluted with the the transversity distribution functions or pretzelosity distribution functions of the transversely polarized proton. We will see that a reasonably description of COMPASS data is achieved with a Boer-Mulders function of the pion used in Ref.~\cite{Boer:1999mm} and TMD PDFs of the proton calculated in the light-cone SU(6) quark-diquark model. We find however that due to the relatively large statistical uncertainties, no clear trend is observed for either of TSAs. We also integrate the two TSAs over the entire kinematic range, and find that the average value of the the $q_{\mathrm{T}}$-weighted TSA $A_{\mathrm{T}}^{\sin \left(2 \varphi-\varphi_{S}\right) q_{\mathrm{T}}/{M_{\pi}}}$ is measured to be below zero. The obtained magnitude of the asymmetry is in agreement with the model calculations of Ref.~\cite{Sissakian:2010zza} and can be used to study the universality of the nucleon transversity function. The $q_{\mathrm{T}}$-weighted TSA $A_{\mathrm{T}}^{\sin \left(2 \varphi+\varphi_{S}\right) q_{\mathrm{T}}^{3}/{2M_{\pi}M_{P}^{2}}}$, which is related to the nucleon pretzelosity TMD PDF, is found to be compatible with zero.

The remainder of this paper is organized as follows. In Section~\ref{sec_model} the $q_{\mathrm{T}}$ weighted TSA framework is presented. Section~\ref{sec_calc} describes the calculation details. The numerical results and discussions are given in Section~\ref{sec_result}, and a brief summary is given in Section~\ref{sec_sum}.

\section{weighted TSAs}
\label{sec_model}

For DY and SIDIS cross sections, TMD factorization was proven to hold~\cite{Collins:2011zzd}, which allows one to express the cross sections as convolutions of hard-scale dependent TMD PDFs, perturbatively calculable hard-scattering parton cross sections and (for SIDIS) parton fragmentation functions. The hard-scale $Q$ in DY is given by the invariant mass of the lepton pair. At leading order, for the case of a transversely polarized target, the corresponding single-polarized Drell-Yan cross section has the general form~\cite{Arnold:2008kf}:
\begin{equation}
\begin{aligned}\label{dycs}
 \frac{\mathrm{d} \sigma_{\mathrm{DY}}}{\mathrm{d} x_{\pi} \mathrm{d} x_{N} \mathrm{d} q_{\mathrm{T}}^{2} \mathrm{d} \varphi_{\mathrm{S}} \mathrm{d} \cos \theta \mathrm{d} \varphi} \propto & \{(1+\cos ^{2} \theta) F_{\mathrm{U}}^{1}\\ &+ \sin ^{2}
\theta \cos 2 \varphi F_{\mathrm{U}}^{\cos 2 \varphi} \\ &+|\vec{S}_{\mathrm{T}}|[(1+\cos ^{2} \theta) \sin \varphi_{\mathrm{S}} F_{\mathrm{T}}^{\sin \varphi_{\mathrm{S}}}\\
&+\sin ^{2} \theta \sin \left(2 \varphi+\varphi_{\mathrm{S}}\right) F_{\mathrm{T}}^{\sin (2 \varphi+\varphi_{\mathrm{S}})} \\
&+\sin ^{2} \theta \sin \left(2 \varphi-\varphi_{\mathrm{S}}\right) F_{\mathrm{T}}^{\sin (2 \varphi-\varphi_{\mathrm{S}})}]\},
\end{aligned}
\end{equation}
where $F_{X}^{[\mathrm{mod}]}=F_{X}^{[\mathrm{mod}]}\left(x_{\pi}, x_{N}, q_{\mathrm{T}}\right)$ are the structure functions, $\varphi(\theta)$ represents the azimuthal (polar) angle of the lepton momentum in the Collins-Soper frame~(see e.g. Ref.~\cite{Arnold:2008kf}) and $\varphi_{\mathrm{S}}$ the azimuthal angle of the target spin vector in the target rest frame~(see e.g. Ref.~\cite{Arnold:2008kf}), respectively.
In the notation $F^{\mathrm{[mod]}}_{X}$ the superscript [mod] indicates the associated modulation, while the subscript X denotes the target polarization states (``U'' stands for unpolarized and ``T'' for transverse polarization). The structure functions can be written as a flavour sum of convolutions of TMD PDFs over the intrinsic momenta of the two colliding quarks $\vec{k}_{a\mathrm{T}}$ and $\vec{k}_{b\mathrm{T}}$.

With the following notation for the convolution of TMD PDFs in the transverse momentum space:
\begin{equation}
\begin{aligned}
\mathcal{C}\left[w\left(\vec{k}_{a\mathrm{T}}, \vec{k}_{b\mathrm{T}}\right) f_{1} \bar{f}_{2}\right] \equiv & \frac{1}{N_{c}} \sum_{q} e_{q}^{2} \int d^{2} \vec{k}_{a\mathrm{T}} d^{2} \vec{k}_{b\mathrm{T}} \\
& \times \delta^{(2)}\left(\vec{q}_{\mathrm{T}}-\vec{k}_{a\mathrm{T}}-\vec{k}_{b\mathrm{T}}\right) w\left(\vec{k}_{a\mathrm{T}}, \vec{k}_{b\mathrm{T}}\right) \\
& \times\left[f_{1}^{q}\left(x_{a}, \vec{k}_{a\mathrm{T}}^{2}\right) f_{2}^{\bar{q}}\left(x_{b}, \vec{k}_{b\mathrm{T}}^{2}\right)\right.\\
&\left.+f_{1}^{\bar{q}}\left(x_{a}, \vec{k}_{a\mathrm{T}}^{2}\right) f_{2}^{q}\left(x_{b}, \vec{k}_{b\mathrm{T}}^{2}\right)\right],
\end{aligned}
\end{equation}
where $N_{c} = 3$ is the number of colours, using the unit vector $\vec{h} \equiv \vec{q}_{\mathrm{T}} / q_{\mathrm{T}}$ one eventually finds the following leading-order structure functions in the CS-frame~\cite{Arnold:2008kf}:
\begin{equation}
F_{\mathrm{U}}^{1}=\mathcal{C}\left[f_{1} \bar{f}_{1}\right],
\end{equation}

\begin{equation}
F_{\mathrm{U}}^{\cos 2 \varphi}=\mathcal{C}\left[\frac{2\left(\vec{h} \cdot \vec{k}_{a\mathrm{T}}\right)\left(\vec{h} \cdot \vec{k}_{b\mathrm{T}}\right)-\vec{k}_{a\mathrm{T}} \cdot \vec{k}_{b\mathrm{T}}}{M_{a} M_{b}} h_{1}^{\perp} \bar{h}_{1}^{\perp}\right],
\end{equation}

\begin{widetext}
\begin{equation}
F_{\mathrm{T}}^{\sin \left(2 \varphi+\varphi_{S}\right)}=-\mathcal{C}\left[\frac{2\left(\vec{h} \cdot \vec{k}_{b\mathrm{T}}\right)\left[2\left(\vec{h} \cdot \vec{k}_{a\mathrm{T}}\right)\left(\vec{h} \cdot \vec{k}_{b\mathrm{T}}\right)-\vec{k}_{a\mathrm{T}} \cdot \vec{k}_{b\mathrm{T}}\right]-\vec{k}_{b\mathrm{T}}^{2}\left(\vec{h} \cdot \vec{k}_{a\mathrm{T}}\right)}{2 M_{a} M_{b}^{2}} h_{1}^{\perp} \bar{h}_{1 T}^{\perp}\right].
\end{equation}
\end{widetext}

\begin{equation}
F_{\mathrm{T}}^{\sin\varphi_{S}}=\mathcal{C}\left[\frac{\vec{h} \cdot \vec{k}_{b\mathrm{T}}}{M_{b}} f_{1} \bar{f}_{1 T}^{\perp}\right],
\end{equation}

\begin{equation}
F_{\mathrm{T}}^{\sin \left(2 \varphi-\varphi_{S}\right)}=-\mathcal{C}\left[\frac{\vec{h} \cdot \vec{k}_{a\mathrm{T}}}{M_{a}} h_{1}^{\perp} \bar{h}_{1}\right].
\end{equation}

In Eq.~(\ref{dycs}) one can identify five different terms, each containing an orthogonal modulation in $\phi$ or $\varphi_{\mathrm{S}}$ and a structure function $F_{\mathrm{X}}^{[\mathrm{mod}]}$.
The standard TSAs are defined as ratios between structure functions, \begin{equation}\label{stdtsa}
A_{\mathrm{U} / \mathrm{T}}^{[\mathrm{mod}]}=F_{\mathrm{U} / \mathrm{T}}^{[\mathrm{mod}]} / F_{\mathrm{U}}^{1}.
\end{equation}
Therefore, measuring TSAs gives access to a ratio of convolutions of TMD PDFs. The convolutions are usually solved assuming a certain $k_{\mathrm{T}}$-dependence of TMD PDFs (e.g. a Gaussian parametrization).

The weighted TSAs have recently received increasing attention because of the de-convolution of TMD PDFs. The weighted TSAs, on the other hand, represent a way to avoid making any assumption on the $k_{\mathrm{T}}$-dependence. They take advantage of the fact that, if one integrates the structure functions over the intrinsic momenta of the two colliding quarks $\vec{k}_{a\mathrm{T}}$ and $\vec{k}_{b\mathrm{T}}$ with appropriate weights $W_{X},$ the convolutions can be easily solved.

The generic $q_{\mathrm{T}}$-weighted TSA can be written as
\begin{equation}\label{wtdtsa}
A_{\mathrm{T}}^{X W_{X}}=\frac{\int \mathrm{d}^{2} \vec{q}_{\mathrm{T}} W_{X} F_{\mathrm{T}}^{X}}{\int \mathrm{d}^{2} \vec{q}_{\mathrm{T}} F_{\mathrm{U}}^{1}}
\end{equation}
When integrating over $\mathbf{q_{\mathrm{T}}}$, the denominator of Eq.~(\ref{wtdtsa}) is easily computed yielding the familiar collinear expression
\begin{equation}
\begin{aligned}\label{fu1}
& \int \mathrm{d}^{2}\mathbf{q_{\mathrm{T}}} F_{\mathrm{U}}^{1} = \int \mathrm{d}^{2}\mathbf{q_{\mathrm{T}}} \mathcal{C}\left[f_{1} \bar{f}_{1}\right] \\
=& \sum_{q} e_{q}^{2}\int \mathrm{d}^{2}\mathbf{q_{\mathrm{T}}} \int \mathrm{d}^{2}\vec{k}_{a\mathrm{T}}  \mathrm{d}^{2}\vec{k}_{b\mathrm{T}} \\
& \quad \times\delta^{(2)}\left(\vec{q}_{\mathrm{T}}-\vec{k}_{a\mathrm{T}}-\vec{k}_{b\mathrm{T}}\right)f_{1}^{q}\left(x_{a}, \vec{k}_{a\mathrm{T}}^{2}\right) f_{1}^{\bar{q}}\left(x_{b}, \vec{k}_{b\mathrm{T}}^{2}\right) \\
=& \sum_{q} e_{q}^{2}\int \mathrm{d}^{2}\vec{k}_{a\mathrm{T}}f_{1}^{q}\left(x_{a}, \vec{k}_{a\mathrm{T}}^{2}\right)  \int\mathrm{d}^{2}\vec{k}_{b\mathrm{T}} f_{1}^{\bar{q}}\left(x_{b}, \vec{k}_{b\mathrm{T}}^{2}\right) \\
=& \sum_{q} e_{q}^{2}f_{1}^{q}\left(x_{a}\right)f_{1}^{\bar{q}}\left(x_{b}\right).
\end{aligned}
\end{equation}
For simplicity, we omit the $Q^{2}$ dependence of parton distributions.
For case $X = \cos \varphi_{S}$, with $W_{X} = \frac{q_{\mathrm{T}}^{2}}{4M_{a}M_{b}}$:
\begin{equation}
\begin{aligned}\label{fu2}
&\int \mathrm{d}^{2}\mathbf{q_{\mathrm{T}}} \frac{q_{\mathrm{T}}^{2}}{4M_{a}M_{b}}F_{\mathrm{U}}^{\cos\varphi_{S}}
= \int \mathrm{d}^{2}\mathbf{q_{\mathrm{T}}}\frac{q_{\mathrm{T}}^{2}}{4M_{a}M_{b}}\mathcal{C}\left[w\left(\vec{k}_{a\mathrm{T}}, \vec{k}_{b\mathrm{T}}\right) h_{1}^{\perp} \bar{h}_{1}^{\perp}\right]  \\
=& \sum_{q} e_{q}^{2}\int \mathrm{d}^{2}\mathbf{q_{\mathrm{T}}} \int \mathrm{d}^{2}\vec{k}_{a\mathrm{T}}  \mathrm{d}^{2}\vec{k}_{b\mathrm{T}} \\ &\quad\times\delta^{(2)}\left(\vec{q}_{\mathrm{T}}-\vec{k}_{a\mathrm{T}}-\vec{k}_{b\mathrm{T}}\right)\frac{q_{\mathrm{T}}^{2}}{4M_{a}M_{b}}\frac{2\vec{k}_{a\mathrm{T}}^{2}\vec{k}_{b\mathrm{T}}^{2}}{q_{\mathrm{T}}^{2}M_{a}M_{b}} \\
&\quad \times h_{1}^{\perp q}\left(x_{a}, \vec{k}_{a\mathrm{T}}^{2}\right)h_{1}^{\perp\bar{q}}\left(x_{b}, \vec{k}_{b\mathrm{T}}^{2}\right) \\
=& 2\sum_{q} e_{q}^{2}\int \mathrm{d}^{2}\vec{k}_{a\mathrm{T}}\frac{\vec{k}_{a\mathrm{T}}^{2}}{2M_{a}^{2}}h_{1}^{\perp q}\left(x_{a}, \vec{k}_{a\mathrm{T}}^{2}\right) \int\mathrm{d}^{2}\vec{k}_{b\mathrm{T}} \frac{\vec{k}_{b\mathrm{T}}^{2}}{2M_{b}^{2}}h_{1}^{\perp\bar{q}}\left(x_{b}, \vec{k}_{b\mathrm{T}}^{2}\right) \\
=& 2\sum_{q} e_{q}^{2}h_{1}^{\perp(1) q}\left(x_{a}\right)h_{1}^{\perp(1)\bar{q}}\left(x_{b}\right).
\end{aligned}
\end{equation}
For case $X = \sin\varphi_{S}$, with $W_{X} = \frac{q_{\mathrm{T}}}{M_{b}}$:
\begin{equation}
\begin{aligned}\label{fut0}
& \int \mathrm{d}^{2}\mathbf{q_{\mathrm{T}}} \frac{q_{\mathrm{T}}}{M_{b}}F_{\mathrm{T}}^{\sin\varphi_{S}}
= -\int \mathrm{d}^{2}\mathbf{q_{\mathrm{T}}} \frac{q_{\mathrm{T}}}{M_{b}}\mathcal{C}\left[\frac{\vec{h} \cdot \vec{k}_{b\mathrm{T}}}{M_{b}} f_{1}\bar{f}_{1T}^{\perp}\right]  \\
=& - \sum_{q} e_{q}^{2}\int \mathrm{d}^{2}\mathbf{q_{\mathrm{T}}} \int \mathrm{d}^{2}\vec{k}_{a\mathrm{T}}  \mathrm{d}^{2}\vec{k}_{b\mathrm{T}} \\ &\quad \times \delta^{(2)}\left(\vec{q}_{\mathrm{T}}-\vec{k}_{a\mathrm{T}}-\vec{k}_{b\mathrm{T}}\right)\frac{q_{\mathrm{T}}}{M_{b}}\frac{\vec{k}_{b\mathrm{T}}^{2}}{q_{\mathrm{T}}M_{b}}f_{1}^{q}\left(x_{a}, \vec{k}_{a\mathrm{T}}^{2}\right) f_{1\mathrm{T}}^{\perp\bar{q}}\left(x_{b}, \vec{k}_{b\mathrm{T}}^{2}\right) \\
=& -2\sum_{q} e_{q}^{2}\int \mathrm{d}^{2}\vec{k}_{a\mathrm{T}}f_{1}^{q}\left(x_{a}, \vec{k}_{a\mathrm{T}}^{2}\right) \int\mathrm{d}^{2}\vec{k}_{b\mathrm{T}}\frac{\vec{k}_{b\mathrm{T}}^{2}}{2M_{b}^{2}} h_{1\mathrm{T}}^{\perp\bar{q}}\left(x_{b}, \vec{k}_{b\mathrm{T}}^{2}\right) \\
=& -2\sum_{q} e_{q}^{2}f_{1}^{q}\left(x_{a}\right)f_{1\mathrm{T}}^{\perp(1)\bar{q}}\left(x_{b}\right).
\end{aligned}
\end{equation}
For case $X = \sin(2\varphi-\varphi_{S})$, with $W_{X} = \frac{q_{\mathrm{T}}}{M_{a}}$:
\begin{equation}
\begin{aligned}\label{fut1}
& \int \mathrm{d}^{2}\mathbf{q_{\mathrm{T}}} \frac{q_{\mathrm{T}}}{M_{a}}F_{\mathrm{T}}^{\sin(2\varphi-\varphi_{S})}
= -\int \mathrm{d}^{2}\mathbf{q_{\mathrm{T}}} \frac{q_{\mathrm{T}}}{M_{a}}\mathcal{C}\left[\frac{\vec{h} \cdot \vec{k}_{a\mathrm{T}}}{M_{a}} h_{1}^{\perp} \bar{h}_{1}\right]  \\
=& - \sum_{q} e_{q}^{2}\int \mathrm{d}^{2}\mathbf{q_{\mathrm{T}}} \int \mathrm{d}^{2}\vec{k}_{a\mathrm{T}}  \mathrm{d}^{2}\vec{k}_{b\mathrm{T}} \\
& \quad \times\delta^{(2)}\left(\vec{q}_{\mathrm{T}}-\vec{k}_{a\mathrm{T}}-\vec{k}_{b\mathrm{T}}\right)\frac{q_{\mathrm{T}}}{M_{a}}\frac{\vec{k}_{a\mathrm{T}}^{2}}{q_{\mathrm{T}}M_{a}}h_{1}^{\perp q}\left(x_{a}, \vec{k}_{a\mathrm{T}}^{2}\right) h_{1}^{\bar{q}}\left(x_{b}, \vec{k}_{b\mathrm{T}}^{2}\right) \\
=& -2\sum_{q} e_{q}^{2}\int \mathrm{d}^{2}\vec{k}_{a\mathrm{T}}\frac{\vec{k}_{a\mathrm{T}}^{2}}{2M_{a}^{2}}h_{1}^{\perp q}\left(x_{a}, \vec{k}_{a\mathrm{T}}^{2}\right) \int\mathrm{d}^{2}\vec{k}_{b\mathrm{T}} h_{1}^{\bar{q}}\left(x_{b}, \vec{k}_{b\mathrm{T}}^{2}\right) \\
=& -2\sum_{q} e_{q}^{2}h_{1}^{\perp(1) q}\left(x_{a}\right)h_{1}^{\bar{q}}\left(x_{b}\right).
\end{aligned}
\end{equation}
A similar integration over $\mathbf{q_{\mathrm{T}}}$ can be applied to the case $X = \sin(2\varphi+\varphi_{S})$, with $W_{X} = \frac{q_{\mathrm{T}}^{3}}{2M_{a}M_{b}^{2}}$, though it is a little more
complicated. In this way the convolution in the numerator can now be carried out and the final expression is:
\begin{equation}
\begin{aligned}\label{fut2}
&\int \mathrm{d}^{2}\mathbf{q_{\mathrm{T}}} \frac{q_{\mathrm{T}}^{3}}{2M_{a}M_{b}^{2}}F_{\mathrm{T}}^{\sin(2\varphi+\varphi_{S})} \\
=& -\int \mathrm{d}^{2}\mathbf{q_{\mathrm{T}}}\frac{q_{\mathrm{T}}^{3}}{2M_{a}M_{b}^{2}}\mathcal{C}\left[w\left(\vec{k}_{a\mathrm{T}}, \vec{k}_{b\mathrm{T}}\right) h_{1}^{\perp} \bar{h}_{1T}^{\perp}\right]  \\
=&- \sum_{q} e_{q}^{2}\int \mathrm{d}^{2}\mathbf{q_{\mathrm{T}}} \int \mathrm{d}^{2}\vec{k}_{a\mathrm{T}}  \mathrm{d}^{2}\vec{k}_{b\mathrm{T}} \delta^{(2)}\left(\vec{q}_{\mathrm{T}}-\vec{k}_{a\mathrm{T}}-\vec{k}_{b\mathrm{T}}\right) \\
&\quad \times\frac{q_{\mathrm{T}}^{3}}{2M_{a}M_{b}^{2}}\frac{\vec{k}_{a\mathrm{T}}^{2}\vec{k}_{b\mathrm{T}}^{4}}{2q_{\mathrm{T}}^{3}M_{a}M_{b}^{2}}h_{1}^{\perp q}\left(x_{a}, \vec{k}_{a\mathrm{T}}^{2}\right) h_{1\mathrm{T}}^{\perp\bar{q}}\left(x_{b}, \vec{k}_{b\mathrm{T}}^{2}\right) \\
=& -2\sum_{q} e_{q}^{2}\int \mathrm{d}^{2}\vec{k}_{a\mathrm{T}}\frac{\vec{k}_{a\mathrm{T}}^{2}}{2M_{a}^{2}}h_{1}^{\perp q}\left(x_{a}, \vec{k}_{a\mathrm{T}}^{2}\right) \\
&\quad \times \int\mathrm{d}^{2}\vec{k}_{b\mathrm{T}} \left(\frac{\vec{k}_{b\mathrm{T}}^{2}}{2M_{b}^{2}}\right)^{2} h_{1\mathrm{T}}^{\perp\bar{q}}\left(x_{b}, \vec{k}_{b\mathrm{T}}^{2}\right) \\
=& -2\sum_{q} e_{q}^{2}h_{1}^{\perp(1) q}\left(x_{a}\right)h_{1\mathrm{T}}^{\perp(2)\bar{q}}\left(x_{b}\right).
\end{aligned}
\end{equation}
In this notation, $f^{(n)}$ or $h^{(n)}$ are the $n$-th $k_{\mathrm{T}}^{2}$-moments of TMD PDFs,
\begin{equation}\label{moment}
h^{(n)q}(x)=\int d^{2} \mathbf{k_{\mathrm{T}}} \left(\frac{k_{\mathrm{T}}^{2}}{2 M^{2}}\right)^{n} h^{q}\left(x, k_{\mathrm{T}}^{2}\right).
\end{equation}

We see from Eqs.~(\ref{wtdtsa}$-$\ref{moment}) that the measurement of the $q_{\mathrm{T}}$ weighted asymmetry: $A_{\mathrm{U}}^{\frac{q_{\mathrm{T}}^{2}}{4M_{a}M_{b}}\cos \varphi_{S}}$ gives access to the first moment of Boer-Mulders functions of the incoming hadrons, $A_{\mathrm{T}}^{\frac{q_{\mathrm{T}}}{M_{b}}\sin \phi_{S}}$ to the first moment of Sivers function of the target nucleon, $A_{\mathrm{T}}^{\frac{q_{\mathrm{T}}^{3}}{4M_{a}M_{b}}\sin \left(2 \phi+\phi_{S}\right)}$ to the first moment of Boer-Mulders function of the beam hadron and to $h_{1\mathrm{T}}^{\perp(2)},$ the second moment of pretzelosity function of the target nucleon, $A_{\mathrm{T}}^{\frac{q_{\mathrm{T}}}{M_{a}}\sin \left(2 \phi-\phi_{S}\right)}$ to the Boer-Mulders function of the beam hadron and $h_{1},$ the transversity function of the target nucleon.

In the case of the pion-induced Drell-Yan processes on a transversely polarized proton target, the three $q_{\mathrm{T}}$-weighted TSAs accessible are
\begin{equation*}
\begin{aligned}
& A_{\mathrm{T}}^{\sin \varphi_{S} \frac{q_{\mathrm{T}}}{M_{\mathrm{p}}}}\left(x_{\pi}, x_{N}\right) \\
 = & -2 \frac{\sum_{q} e_{q}^{2}\left[f_{1, \pi^{-}}^{\bar{q}}\left(x_{\pi}\right) f_{1 \mathrm{T}, \mathrm{p}}^{\perp(1) q}\left(x_{N}\right)+(q \leftrightarrow \bar{q})\right]}{\sum_{q} e_{q}^{2}\left[f_{1, \pi^{-}}^{\bar{q}}\left(x_{\pi}\right) f_{1, \mathrm{p}}^{q}\left(x_{N}\right)+(q \leftrightarrow \bar{q})\right]},
\end{aligned}
\end{equation*}
\begin{equation*}
\begin{aligned}
& A_{\mathrm{T}}^{\sin \left(2 \varphi-\varphi_{\mathrm{S}}\right) \frac{q_{\mathrm{T}}}{M_{\pi}}}\left(x_{\pi}, x_{N}\right) \\
 = & -2 \frac{\sum_{q} e_{q}^{2}\left[h_{1, \pi^{-}}^{\perp(1) \bar{q}}\left(x_{\pi}\right) h_{1, \mathrm{p}}^{q}\left(x_{N}\right)+(q \leftrightarrow \bar{q})\right]}{\sum_{q} e_{q}^{2}\left[f_{1, \pi^{-}}^{\bar{q}}\left(x_{\pi}\right) f_{1, \mathrm{p}}^{q}\left(x_{N}\right)+(q \leftrightarrow \bar{q})\right]},
\end{aligned}
\end{equation*}

\begin{equation*}
\begin{aligned}
& A_{\mathrm{T}}^{\sin \left(2 \varphi+\varphi_{\mathrm{S}}\right) \frac{q_{\mathrm{T}}^{3}}{2 M_{\pi} M_{\mathrm{p}}^{2}}}{\left(x_{\pi}, x_{N}\right)} \\
 =&-2 \frac{\sum_{q} e_{q}^{2}\left[h_{1, \pi^{-}}^{\perp(1) \bar{q}}\left(x_{\pi}\right) h_{1 \mathrm{T}, \mathrm{p}}^{\perp(2) q}\left(x_{N}\right)+(q \leftrightarrow \bar{q})\right]}{\sum_{q} e_{q}^{2}\left[f_{1, \pi^{-}}^{\bar{q}}\left(x_{\pi}\right) f_{1, \mathrm{p}}^{q}\left(x_{N}\right)+(q \leftrightarrow \bar{q})\right]},
\end{aligned}
\end{equation*}
where the sums run over all quarks and antiquarks flavours $q$ with fractional electric charge $e_{q}$, and $M_{\pi}$, $M_{\mathrm{p}}$ represent the pion and proton masses, respectively.

\section{Calculation}
\label{sec_calc}

The measurement of transverse-spin-dependent azimuthal asymmetries in the pion-induced Drell-Yan process, when weighted with the corresponding virtual photon transverse momentum, $q_{\mathrm{T}}$, allows for the extraction of important transverse-momentum-dependent distribution functions. In this paper, we will present a phenomenological analysis of these $q_{\mathrm{T}}$ weighted TSAs in the pion-proton Drell-Yan process contributed by various leading-twist chiral-odd distribution functions. Analogy to Ref.~\cite{Alexeev:2018zvl} and according to the previous analysis, we can see that using the weighted TSAs one can also access the pion Boer-Mulders TMD PDF. It can be independently obtained from COMPASS results using the $A_{\mathrm{T}}^{\sin \left(2 \phi-\varphi_{\mathrm{S}}\right) \frac{q_{\mathrm{T}}}{M_{\pi}}}$ asymmetry:
\begin{equation}
\begin{aligned}\label{w1}
& A_{\mathrm{T}}^{\sin \left(2 \phi-\phi_{S}\right) \frac{q_{\mathrm{T}}}{M_{\pi}}}\left(x_{\pi}, x_{N}\right) \\
=& -2 \frac{\sum_{q} e_{q}^{2}\left[h_{1, \pi}^{\perp(1) \bar{q}}\left(x_{\pi}\right) h_{1, \mathrm{p}}^{q}\left(x_{N}\right)+(q \leftrightarrow \bar{q})\right]}{\sum_{q} e_{q}^{2}\left[f_{1, \pi}^{\bar{q}}\left(x_{\pi}\right) f_{1, \mathrm{p}}^{q}\left(x_{N}\right)+(q \leftrightarrow \bar{q})\right]} \\
\approx & -2 \frac{ h_{1, \pi}^{\perp(1) \overline{\mathrm{u}}}\left(x_{\pi}\right) h_{1, \mathrm{p}}^{\mathrm{u}}\left(x_{N}\right)}{f_{1, \pi}^{\bar{u}}\left(x_{\pi}\right) f_{1, \mathrm{p}}^{u}\left(x_{N}\right)},
\end{aligned}
\end{equation}
where the Boer-Mulders $h_{1,\pi}^{\perp(1)\bar{q}}$ and transversity PDFs $h_{1,\mathrm{p}}^{q}$ of sea quarks are assumed to be zero and only $u$ quark contributions are considered in the denominator, considering an additional suppression coming from the fractional quark charge~(since $e_{u}^{2}=4e_{d}^{2}$).  Looking at the $x_{N}$ vs $x_{\pi}$ distribution in Ref.~\cite{Aghasyan:2017jop}, one can see that COMPASS covers the valence region of both $p$ and $\pi^{-}$; therefore, this assumption that neglecting squared antiquark and strange quark PDF contributions to proton and taking into account the $u$ quark dominance, is fairly justified.

Based on the same assumption as Eq.~(\ref{w1}), another weighted TSA connected to the pion BM TMD PDF is the  $A_{\mathrm{T}}^{\sin \left(2 \phi+\varphi_{\mathrm{S}}\right) \frac{q_{\mathrm{T}}^{3}}{2M_{\pi}M_{P}^{2}}}$ asymmetry, which can be expressed as:
\begin{equation}
\begin{aligned}\label{w2}
& A_{\mathrm{T}}^{\sin \left(2 \phi+\phi_{S}\right) \frac{q_{\mathrm{T}}^{3}}{2M_{\pi}M_{P}^{2}}}\left(x_{\pi}, x_{N}\right) \\
= & -2 \frac{\sum_{q} e_{q}^{2}\left[h_{1, \pi}^{\perp(1) \bar{q}}\left(x_{\pi}\right)  h_{1 \mathrm{T}, \mathrm{p}}^{\perp(2) q}\left(x_{N}\right)+(q \leftrightarrow \bar{q})\right]}{\sum_{q} e_{q}^{2}\left[f_{1, \pi}^{\bar{q}}\left(x_{\pi}\right) f_{1, \mathrm{p}}^{q}\left(x_{N}\right)+(q \leftrightarrow \bar{q})\right]} \\
\approx & -2 \frac{ h_{1, \pi}^{\perp(1) \overline{\mathrm{u}}}\left(x_{\pi}\right)  h_{1 \mathrm{T}, \mathrm{p}}^{\perp(2) u}\left(x_{N}\right)}{f_{1, \pi}^{\bar{u}}\left(x_{\pi}\right) f_{1, \mathrm{p}}^{u}\left(x_{N}\right)},
\end{aligned}
\end{equation}
where $h_{1 \mathrm{T}, \mathrm{p}}^{\perp(2) q}$ is the second $k_{\mathrm{T}}^{2}$-moment of the pretzelosity for the proton.

\subsection{The pion Boer-Mulders function}
\begin{figure}[tbp]
	\small
	\includegraphics[width=9cm]{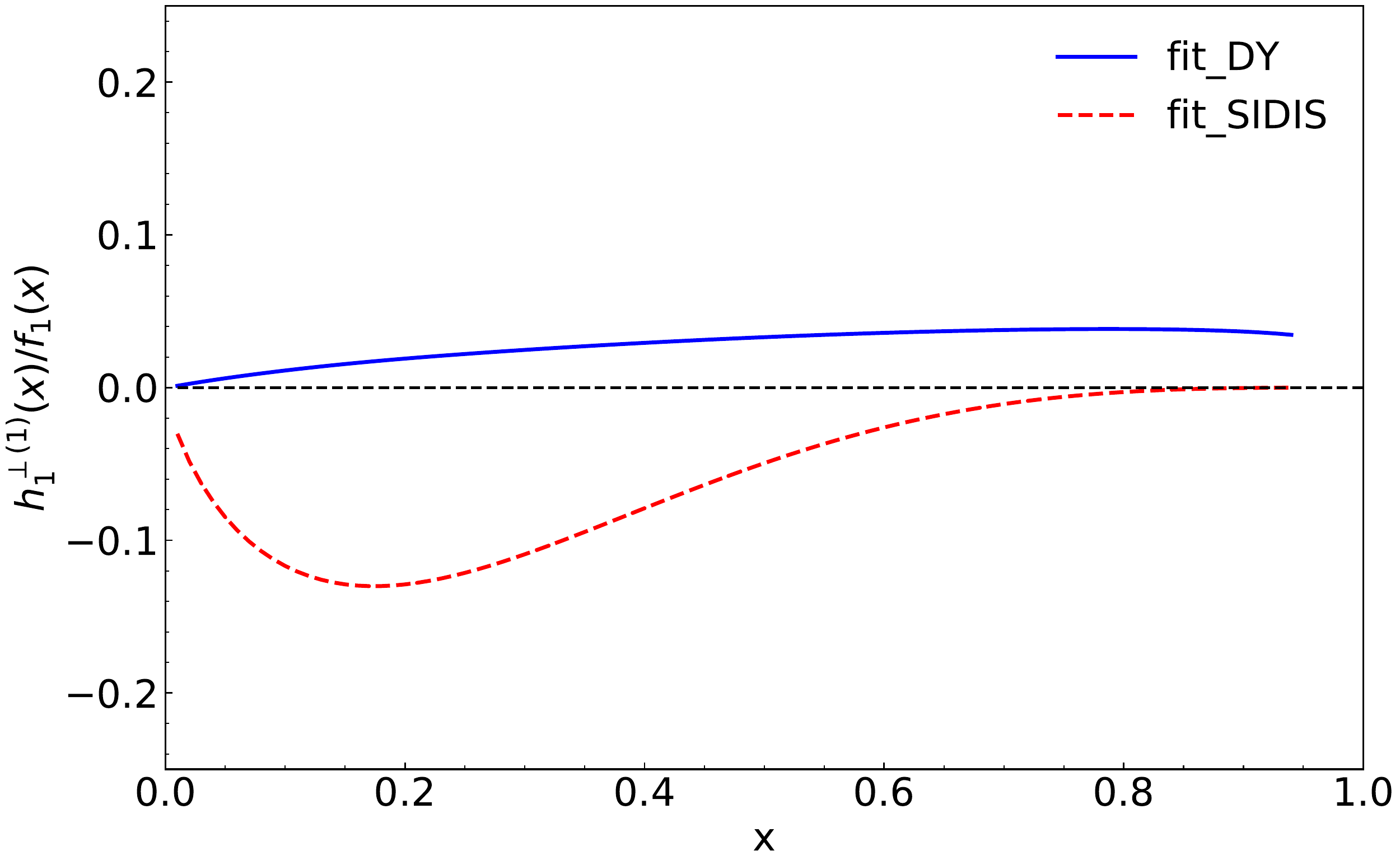}
	\caption{The ratio $h_{1}^{\perp (1)}(x)/f_{1}(x)$ for $u$ quark at $Q^{2} = 25~\mathrm{GeV^{2}}$. The solid blue line corresponds to the DY extraction~\cite{Lu:2009ip}, and the dashed red line correspond to the SIDIS extraction~\cite{Barone:2009hw} of the proton Boer-Mulders function.}
    \label{fig:bm}
\end{figure}
To calculate the asymmetries in Eq.~(\ref{w1}) and Eq.~(\ref{w2}) we need first of all the first moment of pion Boer-Mulders function, $h_{1,\pi}^{\perp(1)\bar{q}}$, which has not been directly obtained from experiments. For the pion Boer-Mulders TMD PDF, $h_{1,\pi}^{\perp q}$, we make use of the model proposed in Ref.~\cite{Boer:1999mm}, which adopts model assumptions on the dependence of the parton distribution function on the quark transverse momentum $k_{\mathrm{T}}$:
\begin{equation}\label{bm}
h_{1}^{\perp a}\left(x, k_{\mathrm{T}}^{2}\right)=\frac{\alpha_{\mathrm{T}}}{\pi} c_{H}^{a} \frac{M_{C} M_{H}}{k_{\mathrm{T}}^{2}+M_{C}^{2}} e^{-\alpha_{\mathrm{T}} k_{\mathrm{T}}^{2}} f_{1}(x),
\end{equation}
with $M_{C}=2.3~\mathrm{GeV}, c_{H}^{a}=1$, $\alpha =1~\mathrm{GeV}^{-1}$, and $M_{H}$ the mass of hadron $H$.

Although, in Ref.~\cite{Boer:1999mm} it was underlined that Eq.~(\ref{bm}) was just a crude model that cannot help to extract the Boer-Mulders function from the data on unpolarized Drell-Yan processes, the model Eq.~(\ref{bm}), as we see below, is sufficient to make a rough prediction on the weighted TSAs. Further more, the first moment of the Boer-Mulders function is calculated according to Eq.~(\ref{moment}) as:
\begin{equation}\label{bm1}
h_{1, \pi}^{\perp(1) \overline{\mathrm{u}}}\left(x_{\pi}\right)=\frac{\alpha_{\mathrm{T}} c_{\pi}^{\overline{\mathrm{u}}} M_{\mathrm{C}}}{M_{\pi}} f_{1, \pi}\left(x_{\pi}\right) \int_{0}^{\infty} \mathrm{d} k_{ \mathrm{T}} \frac{k_{\mathrm{T}}^{3}}{k_{\mathrm{T}}^{2}+M_{\mathrm{C}}^{2}} \mathrm{e}^{-\alpha_{\mathrm{T}} k_{\mathrm{T}}^{2}},
\end{equation}
in which, apparently, the ratio of the first moment of pion BM PDF to the pion PDF is reduced to a constant term. The advantage of the model above is the cancellation in the ratio of the pion PDF, which is still poorly known.

Alternatively, by taking into account the probabilistic interpretation of the $h_{1 u}^{\perp}$ and $f_{1 q}$ as in Ref.~\cite{Sissakian:2010zza}, it is natural to assume that the following relation is satisfied:
\begin{equation}\label{pitop}
\frac{h_{1, \pi^{-}}^{\perp(1)\bar{u}}(x)}{h_{1, p}^{\perp(1)u}(x)}=C_{u} \frac{f_{1, \pi^{-}}^{\bar{u}}(x)}{f_{1, p}^{u}(x)},
\end{equation}
where $C_{u} = \frac{M_{p}c_{\pi}^{u}}{M_{\pi}c_{p}^{u}}$ is consistent with the Boer model~(\ref{bm}). In Ref.~\cite{Sissakian:2010zza},  the value of $C_{u}$ has to be chosen to unity, in order to match the simulation results.  In this way, the pion Boer-Mulders function is connected to the parton distribution functions in the pion and proton, which avoided problems of  calculating the pion PDFs. The similarity of $h_{1}^{\perp(1)}$ for the proton and the pion calculated in Refs.~\cite{Lu:2004hu,Lu:2005rq,Lu:2011qp} also implies that these functions are closely related, because the mechanism that generates them is the same. Hence, a simple way of obtaining the pion Boer-Mulders function, which has not been extracted yet, is to use the value of the proton Boer-Mulders function extracted from other Drell-Yan processes.

So far, the available Boer-Mulders functions for the proton (shown in Fig.~\ref{fig:bm})  were  extracted from unpolarized $pd$ and $pp$ Drell-Yan data measured by the $\mathrm{E866/NuSea}$ Collaboration~\cite{Lu:2009ip} and the $\cos 2\phi$ asymmetry recently measured by the COMPASS and HERMES collaborations in unpolarized semi-inclusive deep inelastic scattering~\cite{Barone:2009hw}. The two parametrizations to calculate the first moment of pion Boer-Mulders function may lead to different results. In this paper, we use the DY type parametrization as adopted in Ref.~\cite{Wang:2018naw} to obtain the proton Boer-Mulders distribution function for making predictions.

Taking the above considerations into account and according to Eq.~(\ref{pitop}), Eqs.~(\ref{w1}) and~(\ref{w2}) are rewritten as:
\begin{equation}\label{w11}
A_{\mathrm{T}}^{\sin \left(2 \varphi-\varphi_{S}\right) \frac{q_{\mathrm{T}}}{M_{\pi}}}\left(x_{\pi}, x_{N}\right)  \approx-2 C_{u}\frac{ h_{1, p}^{\perp(1) u}\left(x_{\pi}\right) h_{1, \mathrm{p}}^{\mathrm{u}}\left(x_{N}\right)}{f_{1, p}^{u}\left(x_{\pi}\right) f_{1, \mathrm{p}}^{u}\left(x_{N}\right)},
\end{equation}
\begin{equation}\label{w21}
A_{\mathrm{T}}^{\sin \left(2 \varphi+\varphi_{S}\right) \frac{q_{\mathrm{T}}^{3}}{2M_{\pi}M_{P}^{2}}}\left(x_{\pi}, x_{N}\right)
\approx-2C_{u} \frac{ h_{1, p}^{\perp(1) u}\left(x_{\pi}\right)  h_{1 \mathrm{T}, \mathrm{p}}^{\perp(2) u}\left(x_{N}\right)}{f_{1, p}^{u}\left(x_{\pi}\right) f_{1, \mathrm{p}}^{u}\left(x_{N}\right)}.
\end{equation}

\subsection{The pretzelosity in the light-cone SU$(6)$ quark-diquak model}

\begin{figure*}[htbp]
	\begin{center}
		\subfigure{\epsfig{figure=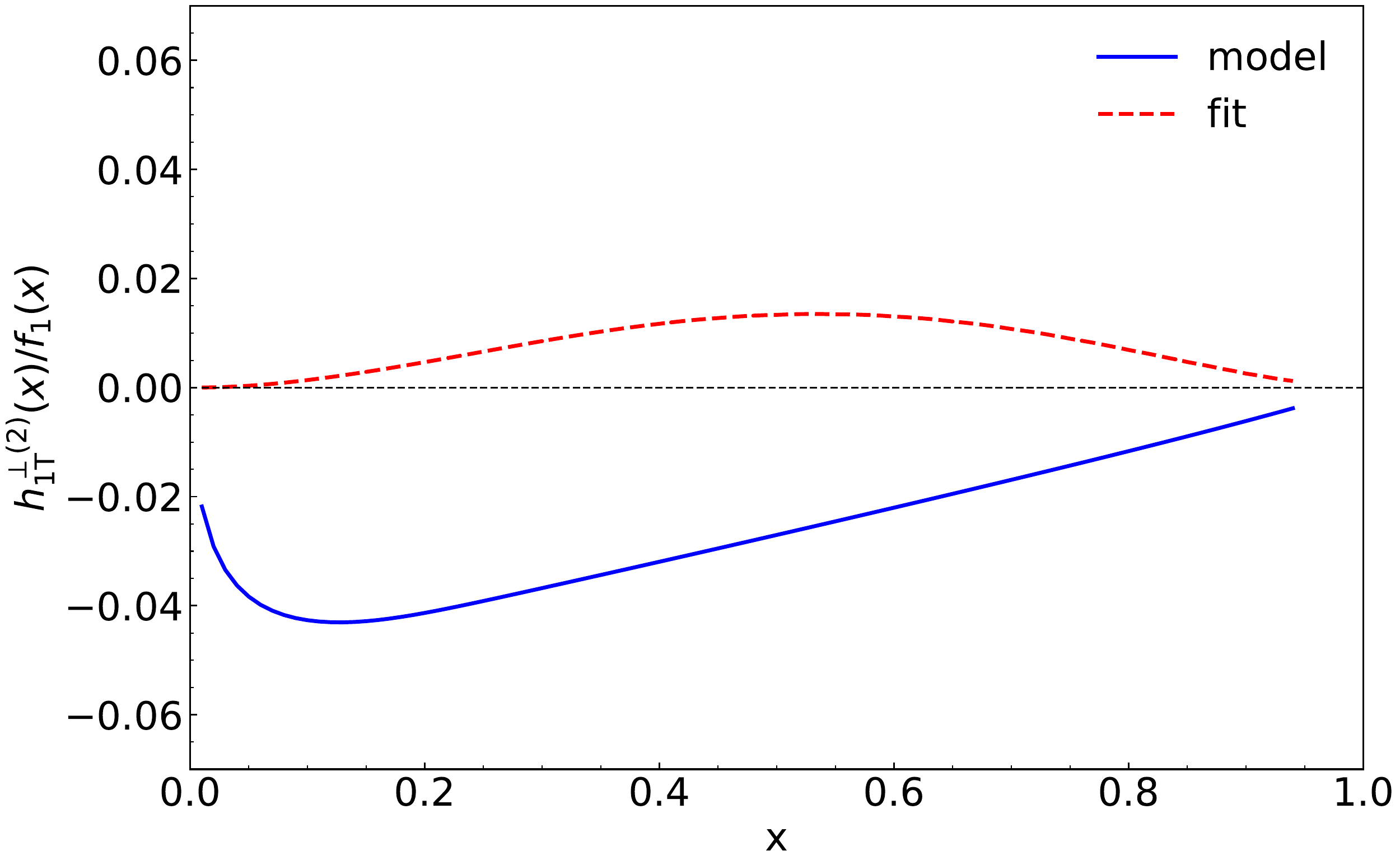, angle=0, width=8.2cm}}
		\subfigure{\epsfig{figure=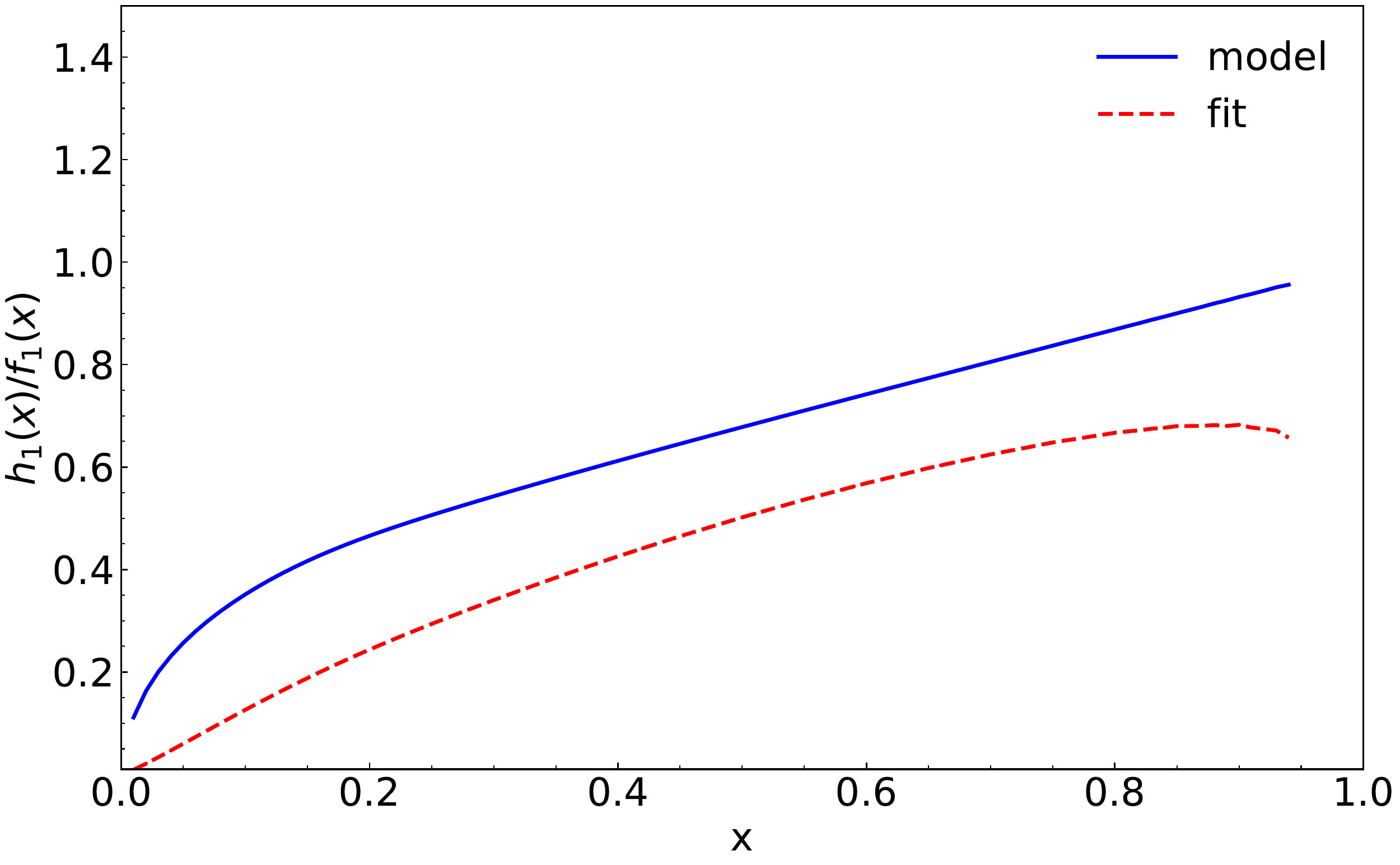, angle=0, width=8.2cm}}
    \end{center}
	\caption{Left panel: The ratio $h_{1\mathrm{T}}^{\perp (2)}(x)/f_{1}(x)$ for $u$ quark at $Q^{2} = 25~\mathrm{GeV^{2}}$. Right panel: The ratio $h_{1}(x)/f_{1}(x)$ for $u$ quark at $Q^{2} = 25~\mathrm{GeV^{2}}$. The solid blue line corresponds to the pretzelosity distribution~\cite{She:2009jq} and the transversity distribution~\cite{Schmidt:1997vm,Ma:1997gy} calculated in the light-cone SU$(6)$ quark-diquark model, and the dashed red line corresponds to the first extraction of the pretzelosity distribution~$(h_{1\mathrm{T}}^{\perp})$~\cite{Lefky:2014eia} and the recent transversity distribution parametrizations~\cite{Kang:2014zza}.}
	\label{fig:pt}
\end{figure*}

 In order to calculate the asymmetries in Eq.~(\ref{w11}) and Eq.~(\ref{w21}) we need first the relevant proton TMD PDFs. If not specified, all the following PDFs refer to those of the proton. The pretzelosity distribution~$(h_{1\mathrm{T}}^{\perp})$, which enters in Eq.~(\ref{w21}), plays a particular role in our understanding of the spin of the nucleon, describing the distribution of a vertically transverse polarized quark in a transversely polarized nucleon~\cite{Mulders:1995dh,Goeke:2005hb,Bacchetta:2006tn}. Model calculations of pretzelosity, including the bag~\cite{Avakian:2008dz,Avakian:2010br} and light-cone quark models~\cite{She:2009jq,Pasquini:2008ax,Boffi:2009sh}, predict negative $u$-quark and positive $d$-quark pretzelosity. However, the first extraction of the pretzelosity distribution~\cite{Lefky:2014eia} from preliminary experimental data on $\sin(3\phi_{h}-\phi_{\mathrm{S}})$ asymmetry, shows tendency for $u$-quark pretzelosity to be positive and $d$-quark pretzelosity to be negative, which are of opposite signs if compared to model calculations. However, the extraction does not give a clear preference on the sign of pretzelosity owing to big errors. Since the pion-induced DY process with a polarized proton is related to the second moment of the pretzelosity distribution $(h_{1\mathrm{T}}^{\perp (2)})$, it opens a way to access the information on the pretzelosity distribution.

In this paper, we calculate the pretzelosity distribution in the light-cone SU$(6)$ quark-diquark model as in Ref.~\cite{She:2009jq}, in which the proton state is constructed by a valence quark and a spectator diquark~\cite{Ma:1996np,Ma:2000cg}, which is a updated version of the quark-diquark model~\cite{Field:1976ve}. The advantage of the light-cone SU$(6)$ quark-diquark model is that some of the gluon and quark effects in the spectator can be effectively described by the diquark with a few parameters. In this model, the Melosh-Wigner rotation~\cite{Melosh:1974cu}, which plays an important role to understand the proton spin puzzle~\cite{Ma:1991xq,Ma:1992sj} due to the relativistic effect of quark transversal motions, has been taken into account. In practice, the light-cone quark-diquark model has been applied to calculate the helicity distributions~\cite{Ma:1996np,Ma:2000cg}, the transversity distributions~\cite{Schmidt:1997vm,Ma:1997gy} and other 3dPDFs~\cite{She:2009jq,Lu:2004au,Zhu:2011zza}, and related azimuthal spin asymmetries in SIDIS processes~\cite{Ma:2001ie,Ma:2002ns}. We therefore expect that the calculation in this paper can make a rough estimate to confront theoretical calculations with the experimental results. \\

The light-cone model results for distributions for $h_{1}$ and $h_{1\mathrm{T}}^{\perp}$ are given as~\cite{Schmidt:1997vm,Ma:1997gy,She:2009jq}:
\begin{equation}\label{trans}
\begin{aligned}
j^{u v}\left(x, k_{\mathrm{T}}^{2}\right)=[&\left.f_{1}^{u v}\left(x, k_{\mathrm{T}}^{2}\right)-\frac{1}{2} f_{1}^{d v}\left(x, k_{\mathrm{T}}^{2}\right)\right] W_{S}^{j}\left(x, k_{\mathrm{T}}^{2}\right) \\
&-\frac{1}{6} f_{1}^{d v}\left(x, k_{\mathrm{T}}^{2}\right) W_{V}^{j}\left(x, k_{\mathrm{T}}^{2}\right), \\
j^{d v}\left(x, k_{\mathrm{T}}^{2}\right)=&-\frac{1}{3} f_{1}^{d v}\left(x, k_{\mathrm{T}}^{2}\right) W_{V}^{j}\left(x, k_{\mathrm{T}}^{2}\right),
\end{aligned}
\end{equation}
where $j=h_{1}, h_{1\mathrm{T}}^{\perp}$, respectively, and the superscript ``$v$'' corresponds to the valence distributions. $W_{S / V}^{j}\left(x, k_{\mathrm{T}}^{2}\right)$ are the rotation factors for the scalar or axial vector spectator-diquark cases. Their explicit form are
\begin{equation}\label{wigner}
\begin{aligned}
W_{D}^{h_{1}}\left(x, k_{\mathrm{T}}^{2}\right)=& \frac{\left(x \mathcal{M}_{D}+m_{q}\right)^{2}}{\left(x \mathcal{M}_{D}+m_{q}\right)^{2}+k_{\mathrm{T}}^{2}}, \\
W_{D}^{h_{1\mathrm{T}}^{\perp}}\left(x, k_{\mathrm{T}}^{2}\right)=&-\frac{2 M_{N}^{2}}{\left(x \mathcal{M}_{D}+m_{q}\right)^{2}+k_{\mathrm{T}}^{2}},
\end{aligned}
\end{equation}
with $\mathcal{M}_{D}^{2}=\frac{m_{q}^{2}+k_{\mathrm{T}}^{2}}{x}+\frac{m_{D}^{2}+k_{\mathrm{T}}^{2}}{1-x}$.

As only valence quark distributions can be directly calculated in this model, one must use other information to take into account contributions from sea and gluon distributions. Now, we will use the phenomenological extraction of the unpolarized distribution functions as an input to calculate the pretzelosity distribution. For example, we will use the CT18 global fit~\cite{Hou:2019efy} parametrization to get $f_{1}(x)$, which have been well tested and constrained by many experiments. And then by combining with some assumed transverse momentum $k_{\mathrm{T}}$ factor one can get phenomenological $f_{1}(x,k_{\mathrm{T}}^{2})$ as input to calculate the second moment of pretzelosity distribution $h_{1\mathrm{T}}^{\perp (2)}(x)$. The calculation results (labelled ``model'') are shown in the left panel of Fig.~\ref{fig:pt}, compared to the pretzelosity distribtion (labelled ``fit'') extracted in~\cite{Lefky:2014eia}.

We adopt the CT18 global fit~\cite{Hou:2019efy} parametrization for the unpolarized distribution in this paper and assume a Gaussian form factor of transverse momentum as suggested in Ref.~\cite{Anselmino:2005nn}:
\begin{equation}
 f_{1}\left(x, \boldsymbol{p}_{\perp}\right)=f_{1}(x) \frac{\exp \left(-p_{\perp}^{2} / p_{a v}^{2}\right)}{\pi p_{a v}^{2}},
 \end{equation}
with $p_{a v}^{2}=0.25~\mathrm{GeV}^{2}$. The parameter $p_{a v}^{2}$ describes the Gaussian width of the transverse momentum distribution.

As stated in Sec.~\ref{sec_model}, the advantage of $q_{\mathrm{T}}$-weighting approach is that it avoids the model dependence of $k_{\mathrm{T}}$, namely Eqs. (\ref{w11}-\ref{w21}). As studied in Ref.~\cite{She:2009jq}, there is a necessity to include transverse momentum dependence of PDFs in data analysis. In the PDF inputs of Eqs. (\ref{w11}-\ref{w21}), the $k_{\mathrm{T}}$ factor included in the SU(6) quark-diquark model we adopted reflects the information of the transverse component of the intrinsic parton momentum inside hadrons. The distribution functions calculated according to the SU(6) quark-diquark model have been integrated over the transverse momentum $k_{\mathrm{T}}$ before entering Eqs.~(\ref{w11}-\ref{w21}), which does not affect the weights. It is worth mentioning that, this is different from assuming a certain $k_{\mathrm{T}}$-dependence of TMD PDFs to solve the convolution mentioned above.

\subsection{The transversity in the light-cone SU$(6)$ quark-diquak model}

As involved in Eq.~(\ref{w11}), the transversity parton distribution function $h_{1}$ is an essential piece of information on the spin structure of quarks in the nucleon at leading twist. The transversity distribution, first introduced by Ralston and Soper in 1979~\cite{Ralston:1979ys} and named by Jaffe and Ji in 1991~\cite{Jaffe:1991kp}, enjoys the same status with the unpolarized distributions, $f_{1}(x)$, and the helicity distributions $g_{1}(x)$. Transversity describes the correlation between the transverse polarization of the nucleon and the transverse polarization of its constituent partons. It measures the difference of the number of quarks with transverse polarization parallel and antiparallel to the proton transverse polarization. Recently, the extraction of the $h_{1}$ receives extensive attentions~\cite{Anselmino:2013vqa,Anselmino:2015sxa,Kang:2015msa,Radici:2015mwa,Radici:2016lam,Radici:2018iag} and is of great significance because of the tensor charge $\delta q$ playing an important role in detecting possible signals of new physics in low-energy experiments~\cite{Dubbers:2011ns,Bhattacharya:2011qm,Cirigliano:2013xha,Bhattacharya:2015esa,Courtoy:2015haa,Tomalak:2017owk,Yamanaka:2017mef}.

Over the past decades, there are several theoretical models about the transversity distribution of the nucleon~(see Ref.~\cite{Barone:2001sp} for a review), such as the MIT bag model~\cite{Chodos:1974je,Chodos:1974pn,DeGrand:1975cf} and the colour dielectric model~\cite{Pirner:1991im,Birse:1991cx,Banerjee:1992jc}, chiral quark soliton model~\cite{Diakonov:1985eg,Diakonov:1987ty} and chiral quark model~\cite{Manohar:1983md}, light-cone models~\cite{Dirac:1949cp,Leutwyler:1977vy}, which based on the Melosh-Wigner rotation~\cite{Wigner:1939cj,Melosh:1974cu,Buccella:1974bz}, and so on. The effect of Melosh-Wigner rotation is also important in leading twist distribution functions such as the transversity distribution, which have been extensively discussed in light-cone formalism in Refs.~\cite{Schmidt:1997vm,Ma:1997gy}. In this paper, we calculate the pretzelosity distribution in the light-cone SU(6) quark-diquark model as in Refs.~\cite{Schmidt:1997vm,Ma:1997gy}. We find that all the distributions have similar form as Eq.~(\ref{trans}) shows, but with different Melosh-Wigner rotation factors as shown in Eq.~(\ref{wigner}). The calculation results (labelled ``model'') are shown in  the right panel of Fig.~\ref{fig:pt}, compared to the most recent extraction of transversity (labelled ``fit'')~\cite{Kang:2014zza}.

\section{Results and Discussion}
\label{sec_result}

\begin{figure*}[htbp]
	\begin{center}
		\subfigure{\epsfig{figure=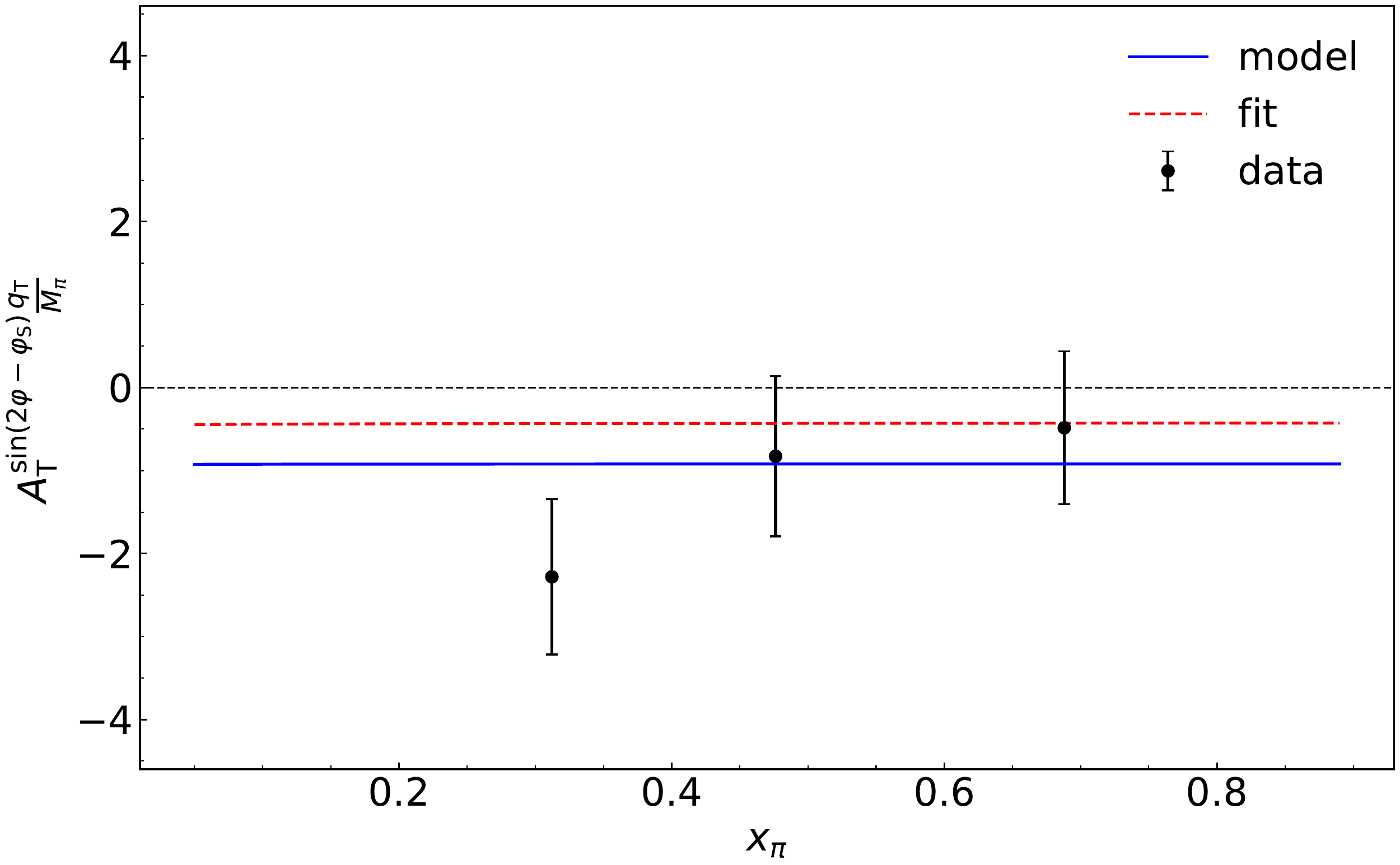, angle=0, width=8.2cm}}
		\subfigure{\epsfig{figure=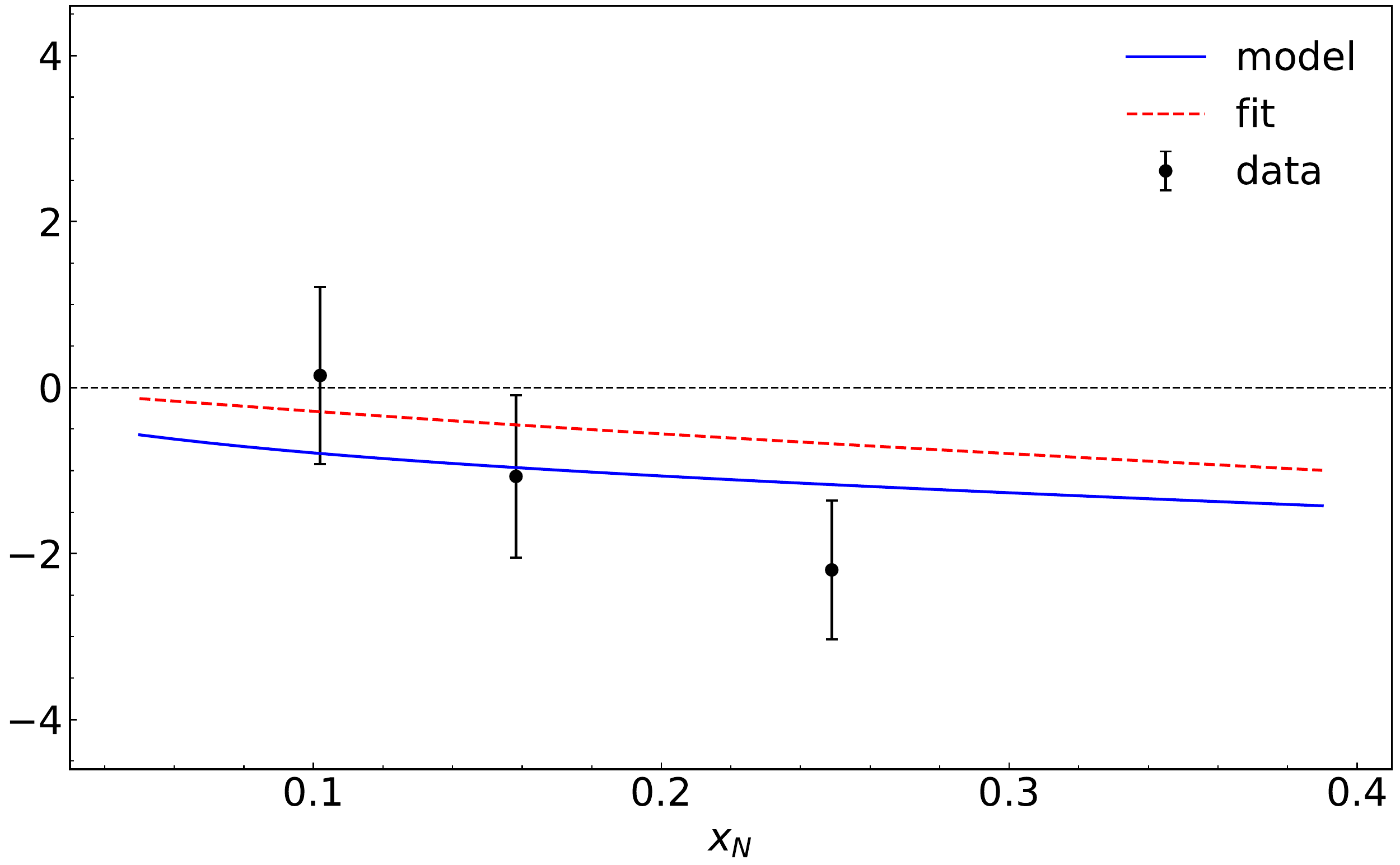, angle=0, width=8.2cm}}
        \subfigure{\epsfig{figure=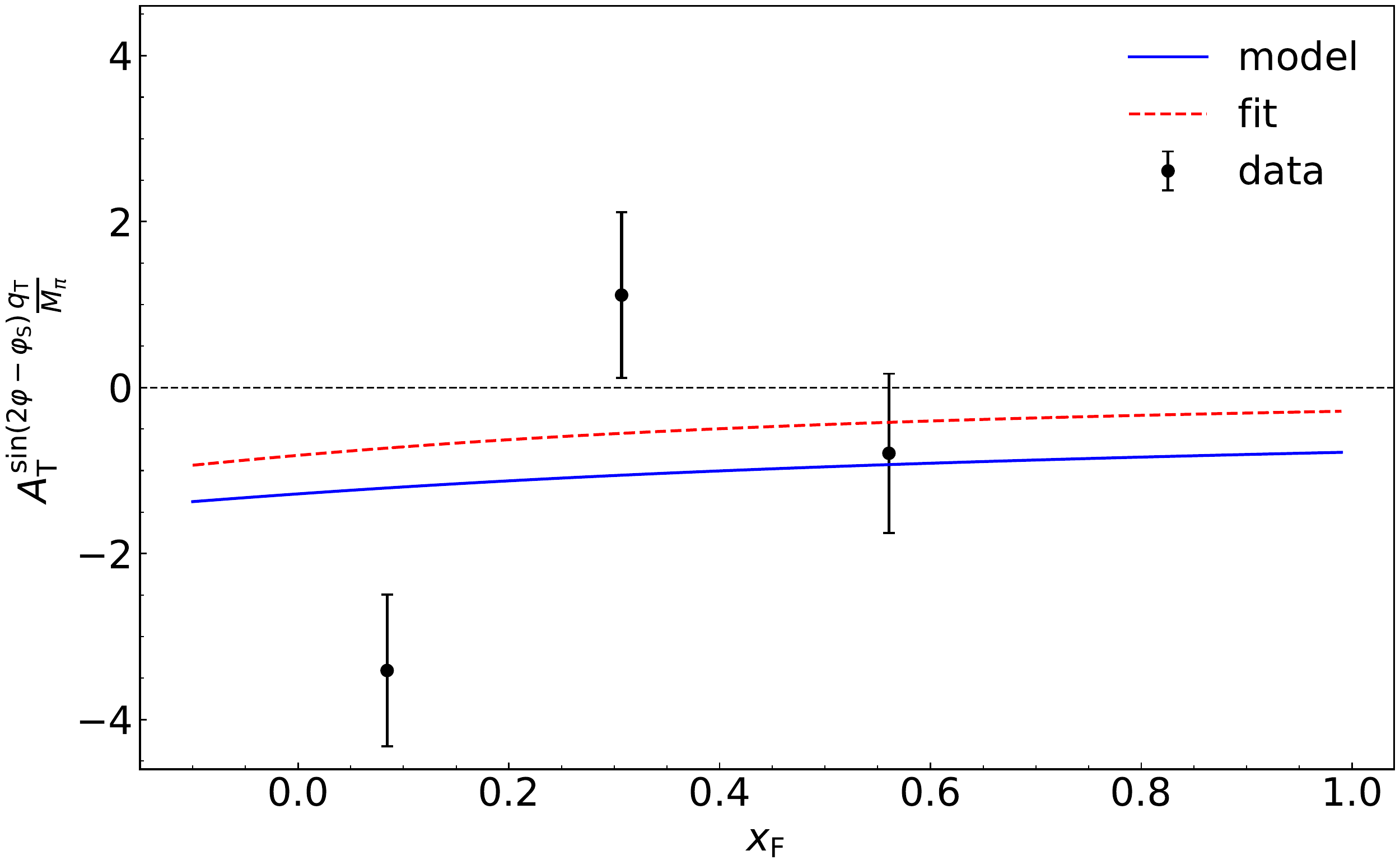, angle=0, width=8.2cm}}
		\subfigure{\epsfig{figure=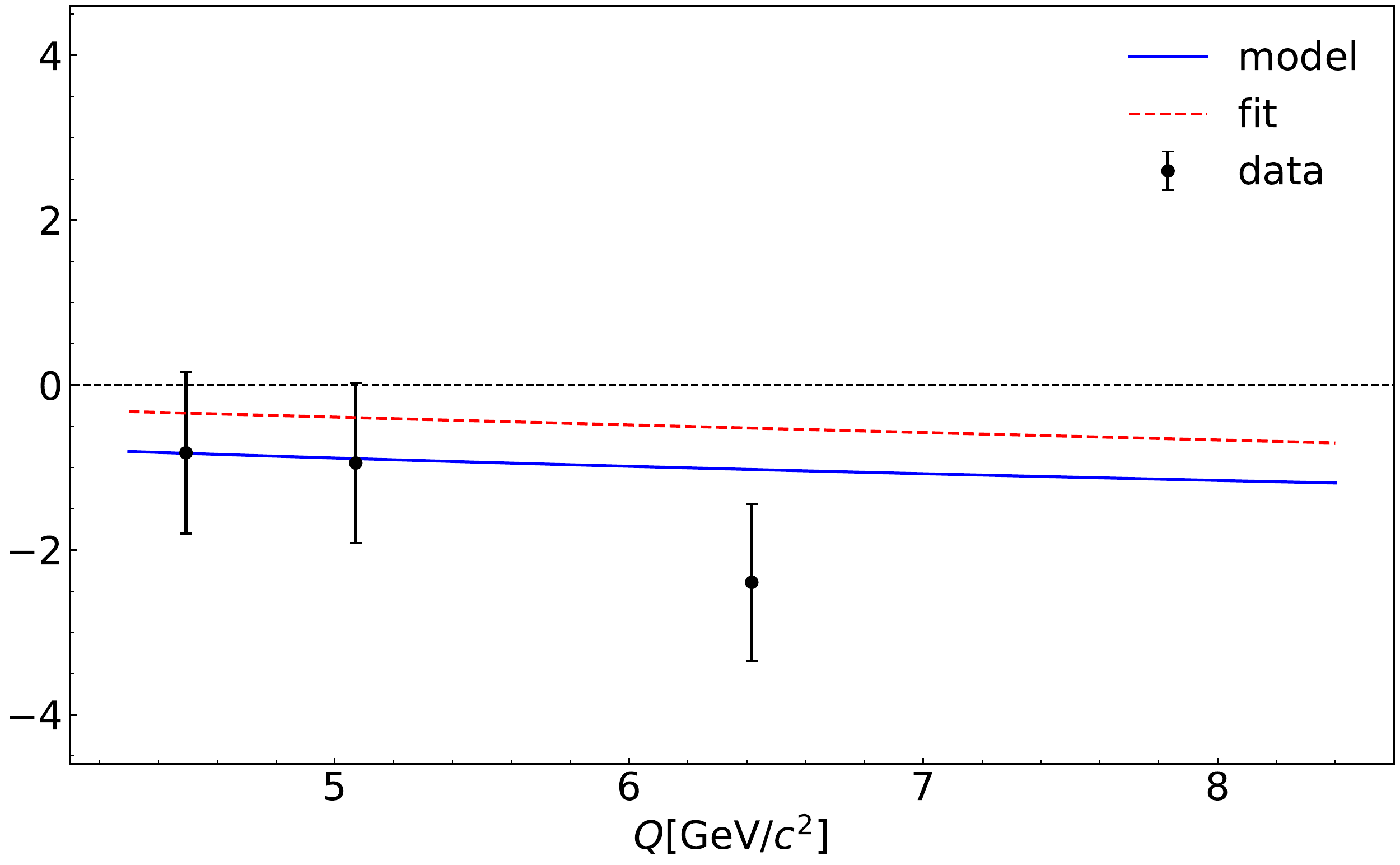, angle=0, width=8.2cm}}
	\end{center}
	\caption{Theoretical calculations and experimental statistical errors on the $q_{\mathrm{T}}$-weighted $\sin\left(2 \varphi-\varphi_{S}\right)$ asymmetries in various kinematic dependence for a DY measurement $\pi^{-}p^{\uparrow}\rightarrow \mu^{+}\mu^{-}X$ with a $190~\mathrm{GeV}/c$ $\pi^{-}$ beam in the high-mass region 4~$\mathrm{GeV}/c^{2}$ $<$ $M_{\mu\mu}$ $<$ 9~$\mathrm{GeV}/c^{2}$~\cite{Longo:2019bih}. Feynman-$x$ or $x_{F}$, is a variable of interest that sheds light on the longitudinal structure of the initial state of the interacting quark. The solid blue line represents the model calculated results, and the dashed red line represents the fitted PDF calculated results.}
	\label{fig:tx}
\end{figure*}

\begin{figure*}[htbp]
	\begin{center}
		\subfigure{\epsfig{figure=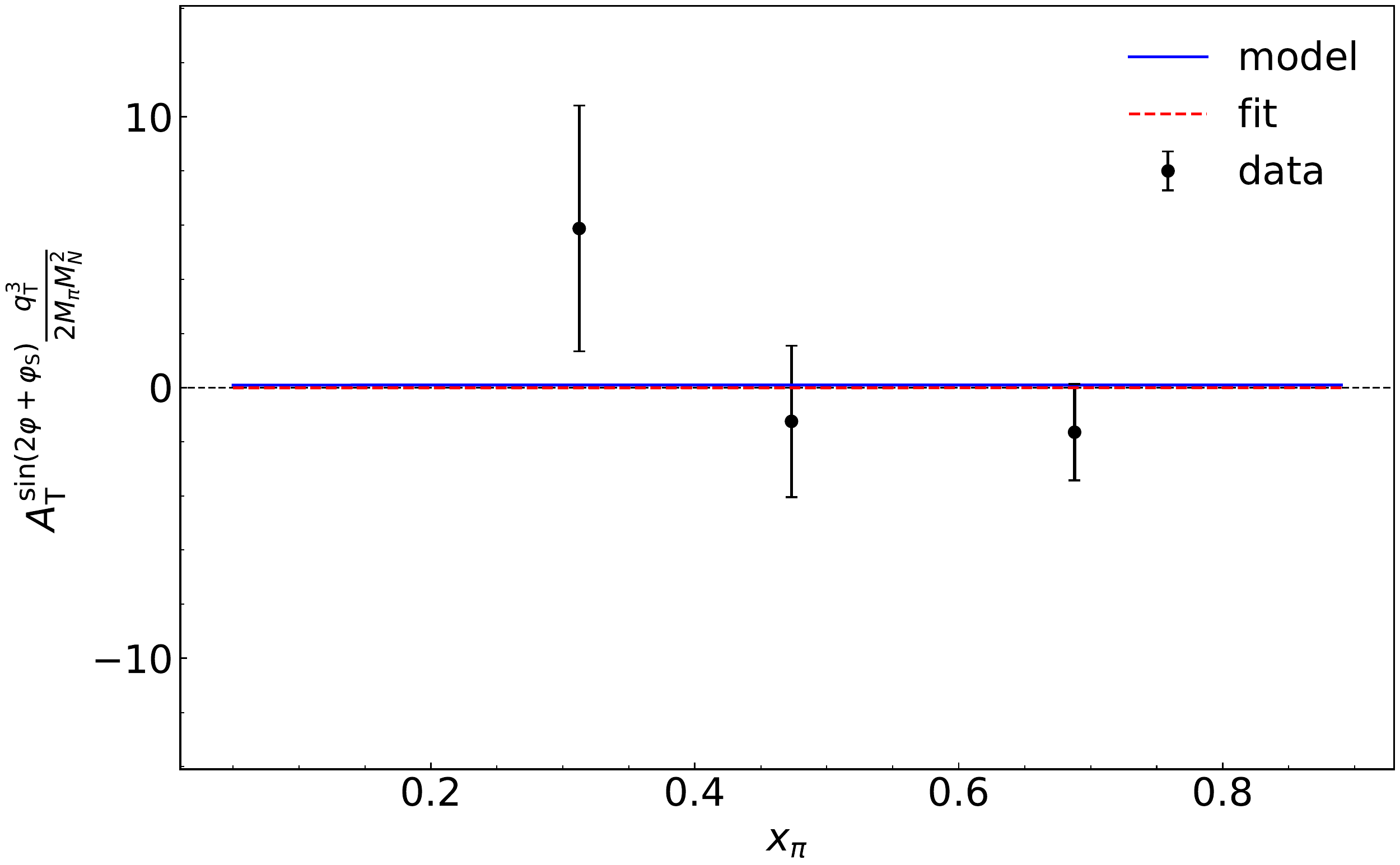, angle=0, width=8.2cm}}
		\subfigure{\epsfig{figure=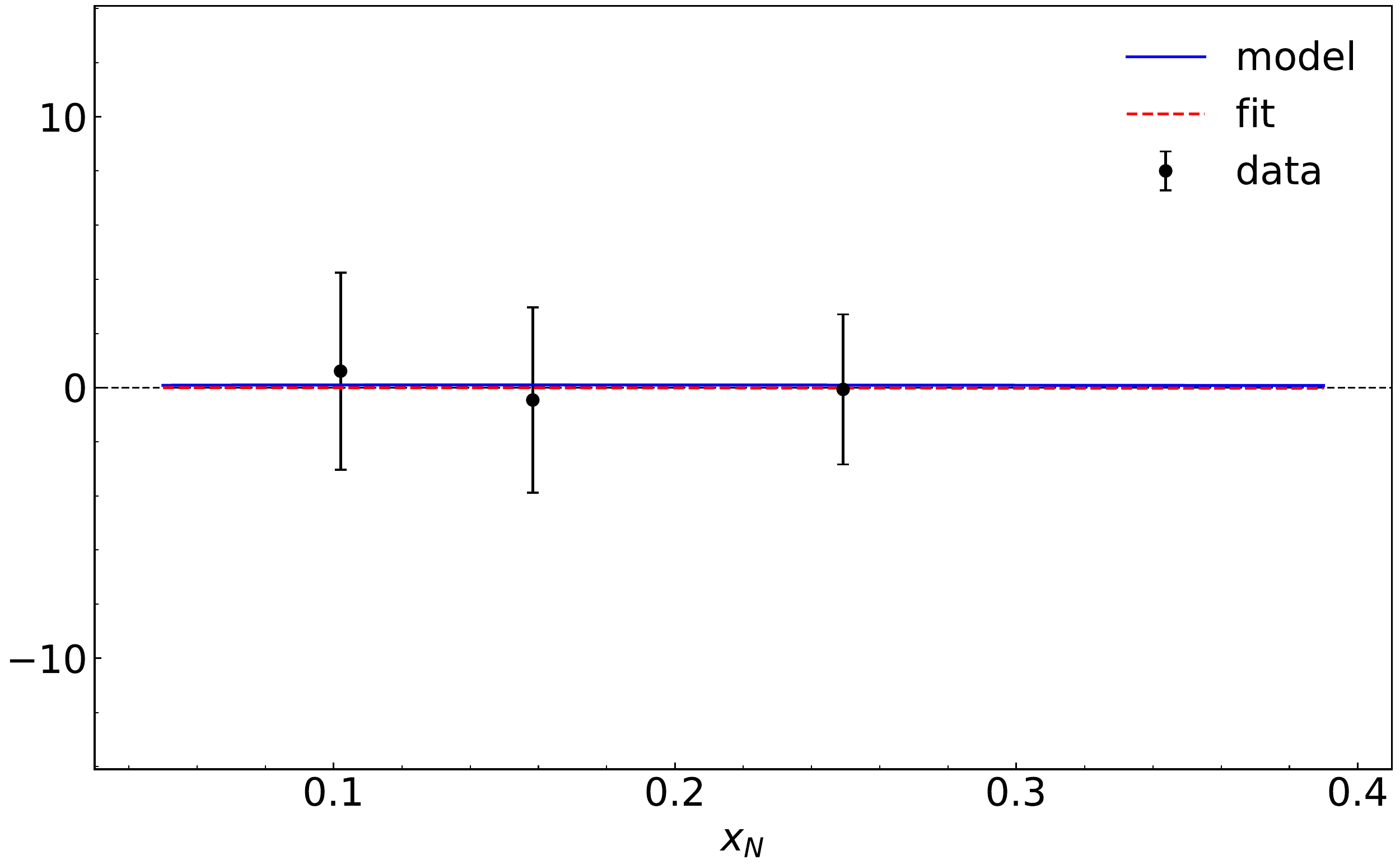, angle=0, width=8.2cm}}
        \subfigure{\epsfig{figure=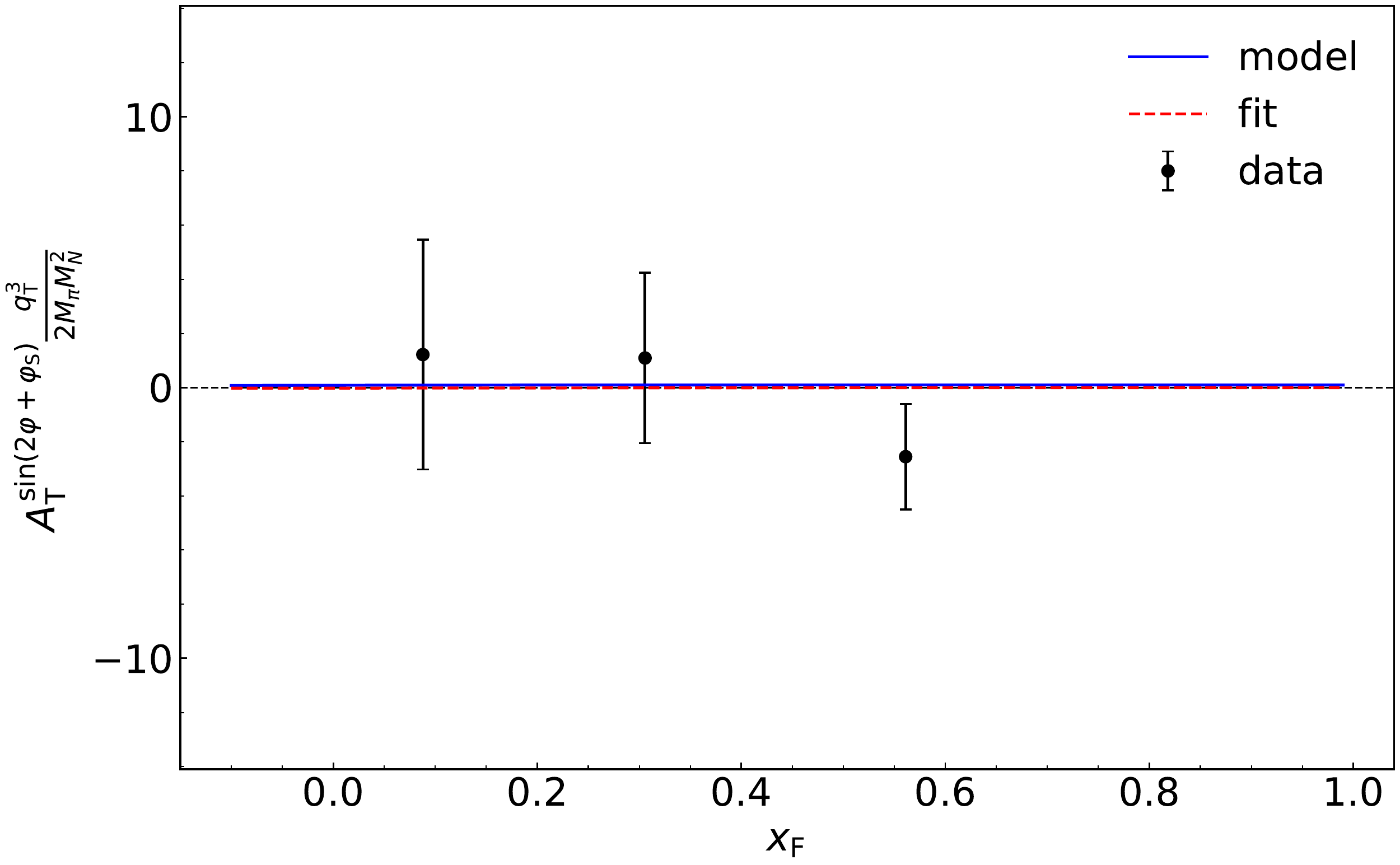, angle=0, width=8.2cm}}
		\subfigure{\epsfig{figure=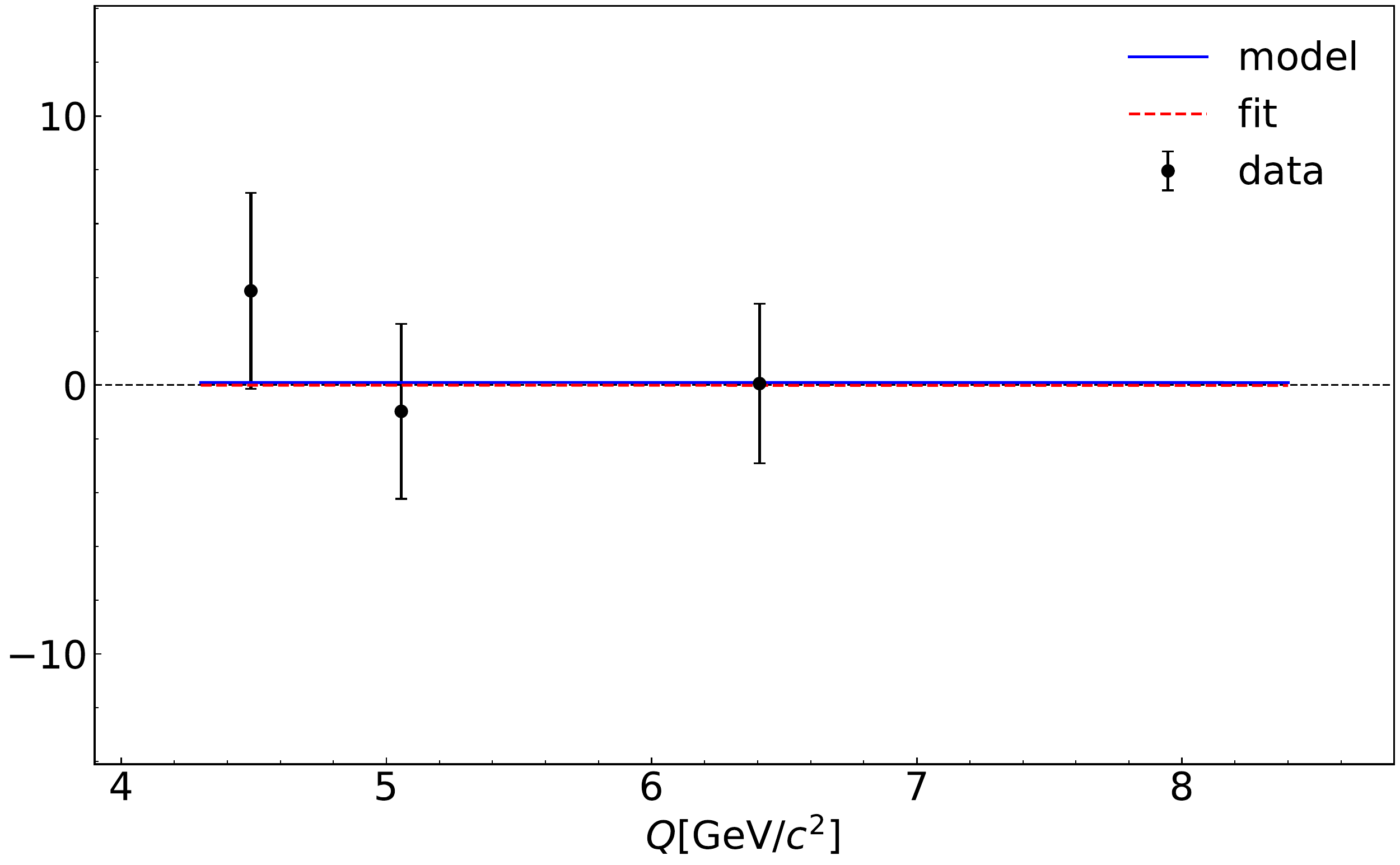, angle=0, width=8.2cm}}
	\end{center}
	\caption{Same as in Fig.~\ref{fig:tx} but for $q_{\mathrm{T}}$-weighted $\sin\left(2 \varphi+\varphi_{S}\right)$ asymmetries.}
	\label{fig:px}
\end{figure*}

\begin{figure*}[htbp]
	\begin{center}
		\subfigure{\epsfig{figure=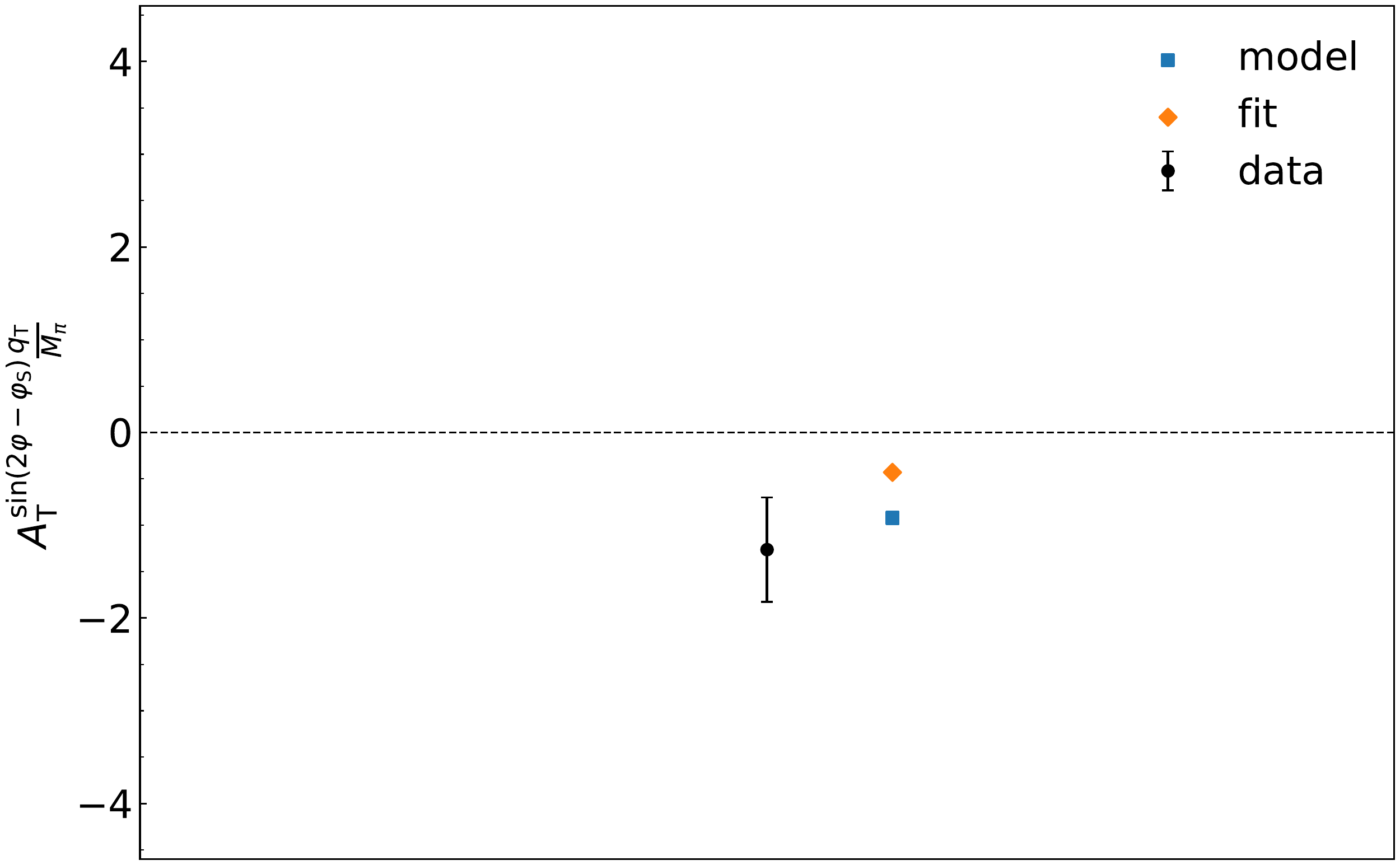, angle=0, width=8.2cm}}
		\subfigure{\epsfig{figure=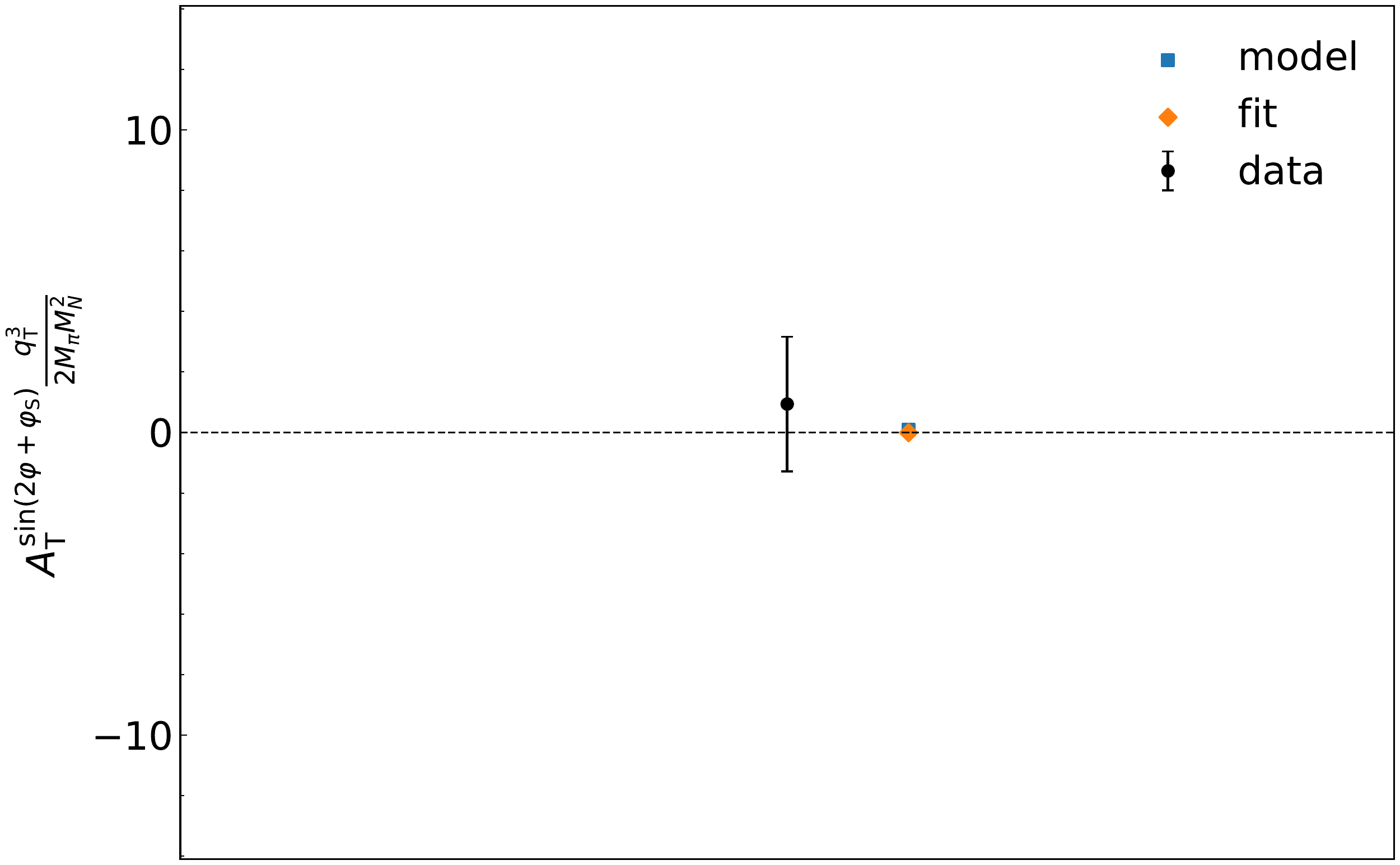, angle=0, width=8.2cm}}
	\end{center}
	\caption{$q_{\mathrm{T}}$-weighted Drell-Yan TSAs integrated over the entire kinematic range. The blue square represents the calculated results, and the red diamond represents the fit results. The data points include the estimated corrections for systematic errors. The error bars contain statistical only.}
	\label{fig:inte}
\end{figure*}

As mentioned above, we adopt the Boer model to relate the Boer-Mulders distribution function of the pion with that of the proton~(Eq.~(\ref{pitop})), with the latter being the parametrization extracted from fitting data~\cite{Lu:2009ip}. The pretzelosity and transversity distribution functions of the proton are adopted from both the light-cone SU(6) quark-diquark model calculations~(Eq.~(\ref{trans})) and parametrizations~\cite{Lefky:2014eia,Kang:2014zza} respectively. Then, we use the DY type parametrization as adopted in Ref.~\cite{Wang:2018naw} to obtain the proton Boer-Mulders distribution function and Eq.~(\ref{trans}) to obtain the proton TMD PDFs. Therefore we are able to calculate the the asymmetries in Eq.~(\ref{w11}) and Eq.~(\ref{w21}). The asymmetries are shown as a function of one variable at a time, $x_{N}$, $x_{\pi}$, $x_{F}$ or $Q$, while always integrating over all the other variables. The integration over the unobserved variables has been performed over the measured ranges of the COMPASS experiment,
\begin{equation*}
\begin{array}{rlr}
0.05<x_{N}<0.4, & 0.05<x_{\pi}<0.4, & 4.3~\mathrm{GeV}<Q<8.5~\mathrm{GeV}, \\
& -0.3<x_{F}<1, & s=357~\mathrm{GeV}^{2},
\end{array}
\end{equation*}
where Feynman-$x$ or $x_{F}$, is a variable of interest that sheds light on the longitudinal structure of the initial state of the interacting quark.

\begin{widetext}
 In Figs.~\ref{fig:tx} and \ref{fig:px} we plot the two transverse spin dependent azimuthal asymmetries $A_{\mathrm{T}}^{\sin \left(2 \varphi-\varphi_{S}\right) q_{\mathrm{T}}/{M_{\pi}}}$ and $A_{\mathrm{T}}^{\sin \left(2 \varphi+\varphi_{S}\right) q_{\mathrm{T}}^{3}/{2M_{\pi}M_{P}^{2}}}$, measured at COMPASS with a transversely polarized $\mathrm{NH_{3}}$ target. As shown in Figs.~\ref{fig:tx} and \ref{fig:px}, the solid blue line represents the model calculated results, and the dashed red line represents the fitted PDF calculated results, respectively. Due to the relatively large statistical uncertainties, no clear trend is observed for either of TSAs.  As we can see, in the case of the Boer-Mulders-transversity asymmetry $A_{\mathrm{T}}^{\sin \left(2 \varphi-\varphi_{S}\right) q_{\mathrm{T}}/{M_{\pi}}}$, shown in Fig.~\ref{fig:tx}, the preliminary results obtained with model calculation are found to be in agreement with the published data~\cite{Longo:2019bih}. In the case of the Boer-Mulders-pretzelosity asymmetry $A_{\mathrm{T}}^{\sin \left(2 \varphi+\varphi_{S}\right) q_{\mathrm{T}}^{3}/{2M_{\pi}M_{P}^{2}}}$, shown in Fig.~\ref{fig:px}, the present transversely polarized DY data on azimuthal asymmetries, although still preliminary, represent a clear manifestation of the Boer-Mulders effect. However, they are not sufficient to allow a clear discrimination on the sign of prezelosity distribution.

The calculation and the data for the two extracted TSAs integrated over the entire kinematic range are shown in Fig.~\ref{fig:inte}. The average value of the the $q_{\mathrm{T}}$-weighted TSA $A_{\mathrm{T}}^{\sin \left(2 \varphi-\varphi_{S}\right) q_{\mathrm{T}}/{M_{\pi}}}$ is measured to be below zero. The obtained magnitude of the asymmetry is in agreement with the model calculations of Ref.~\cite{Sissakian:2010zza} and can be used to study the universality of the nucleon transversity function. The $q_{\mathrm{T}}$-weighted TSA $A_{\mathrm{T}}^{\sin \left(2 \varphi+\varphi_{S}\right) q_{\mathrm{T}}^{3}/{2M_{\pi}M_{P}^{2}}}$, which is related to the nucleon pretzelosity TMD PDF, is found to be compatible with zero. As we discussed above, the sign difference of pretzelosity distribution between model calculations and the first extraction from preliminary experimental data, suggests that the pion-induced DY process with a polarized proton opens a way to access the information on the pretzelosity distribution. High precision experiments are expected to shed light on the spin structure of the nucleon.
\end{widetext}

\section{Summary}
\label{sec_sum}

In summary, we investigate the transverse spin dependent azimuthal asymmetries and confirm the Boer-Mulders effect in TSAs, although still preliminary. With the $u$ quark dominance approximation, the pion Boer-Mulders function obtained by the Boer model can reasonably describe the two asymmetries. Compared with the proton Boer-Mulders function, although process dependent, the pion Boer-Mulders function is with the same sign as that of the proton, offering a hint towards the PDF universality between proton and pion Boer-Mulders functions. Contrast to the Boer-Mulders and Sivers functions, the transversity and pretzelosity distributions are universal and process independent, sharing the same sign in both DY and SIDIS processes. Due to large uncertainties, we are still unable to discriminate the sign of the pretzelosity distribution, and high precision experiments are needed to clarify this point.

\begin{acknowledgements}
This work is supported by the National Natural Science Foundation of China~(Grants No.~12075003).
\end{acknowledgements}



\end{document}